\documentclass[11pt,a4paper]{article}

\usepackage{amssymb}
\usepackage[margin=1in]{geometry}

\newcommand*{\quo}[1]{``{#1}''} 
\newcommand*{\set}[1]{\ensuremath{\{#1\}}} 
\newcommand*{\st}{\mid\,} 
\newcommand*{\card}[1]{\ensuremath{\|{#1}\|}} 
\newcommand*{\abs}[1]{\ensuremath{|{#1}|}} 
\newcommand*{\flor}[1]{\ensuremath{\lfloor{#1}\rfloor}} 
\newcommand*{\ceil}[1]{\ensuremath{\lceil{#1}\rceil}} 
\newcommand*{\str}{^{\ast}} 
\newcommand*{\ini}{_{\scriptscriptstyle\mathrm{I}}} 
\newcommand*{\acc}{_{\scriptscriptstyle\mathrm{A}}} 
\newcommand*{\rej}{_{\scriptscriptstyle\mathrm{R}}} 
\newcommand*{\neu}{_{\scriptscriptstyle\mathrm{N}}} 
\newcommand*{\exs}{_{\scriptscriptstyle\exists}} 
\newcommand*{\uns}{_{\scriptscriptstyle\forall}} 
\newcommand*{\yes}{_{\mathtt{yes}}} 
\newcommand*{\no}{_{\mathtt{no}}} 
\newcommand*{\allone}{{\scriptscriptstyle\forall_1}} 
\newcommand*{\exzero}{{\scriptscriptstyle\exists_0}} 
\newcommand*{\xth}{\textsuperscript{th}} 
\newcommand*{\xst}{\textsuperscript{st}} 
\newcommand*{\rvs}{^{\scriptscriptstyle\mathrm{R}}} 

\newcommand*{\trs}[1]{\ \raisebox{-0.3ex}[0ex]{%
  $\stackrel{#1}{\raisebox{0ex}[0.65ex]{$\longrightarrow$}}$}\ }
\newcommand*{\leftd}{{\scriptscriptstyle\hookleftarrow}} 
\newcommand*{\rightd}{{\scriptscriptstyle\hookrightarrow}} 
\newcommand*{\stayd}{{\scriptscriptstyle\circlearrowright}} 

\newcommand*{\eps}{\ensuremath{\varepsilon}} 
\newcommand{\nnumber}{\mbox{$\mathcal{N}_{\scriptscriptstyle\!+}$}} 
\newcommand{\mba}{\mathbb{A}} 
\newcommand{\mbr}{\mathbb{R}} 
\newcommand{\mbn}{\mathbb{N}} 
\newcommand*{\lcm}{\mathop{\mathrm{lcm}}\nolimits} 
\newcommand*{\iit}[1]{\emph{\/(#1)\/}} 
\newcommand*{\lbr}{\linebreak[0]} 

\newcommand*{\cld}{\ensuremath{\mathcal{D}}}
\newcommand*{\cln}{\ensuremath{\mathcal{N}}}
\newcommand*{\cla}{\ensuremath{\mathcal{A}}}
\newcommand*{\clp}{\ensuremath{\mathcal{P}}}
\newcommand*{\clu}{\ensuremath{\mathcal{U}}}

\newcommand*{\dfa}{\textsc{dfa}}
\newcommand*{\dfas}{\textsc{dfa}s}
\newcommand*{\svfa}{\textsc{svfa}}
\newcommand*{\svfas}{\textsc{svfa}s}
\newcommand*{\nfa}{\textsc{nfa}}
\newcommand*{\nfas}{\textsc{nfa}s}
\newcommand*{\afa}{\textsc{afa}}
\newcommand*{\afas}{\textsc{afa}s}
\newcommand*{\pfa}{\textsc{pfa}}
\newcommand*{\pfas}{\textsc{pfa}s}
\newcommand*{\qfa}{\textsc{qfa}}
\newcommand*{\qfas}{\textsc{qfa}s}
\newcommand*{\obbd}{\textsc{obbd}}
\newcommand*{\owe}{\textrm{\upshape\scriptsize 1}${}_\varepsilon$} 
\newcommand*{\ow}{\textrm{\upshape\scriptsize 1}} 
\newcommand*{\tw}{\textrm{\upshape\scriptsize 2}} 
\newcommand*{\rt}{\textrm{\upshape\footnotesize rt}} 

\newcommand*{\ttp}{\ensuremath{\mathtt{P}}}
\newcommand*{\ttr}{\ensuremath{\mathtt{R}}}
\newcommand*{\ttrc}{\ensuremath{\overline{\mathtt{R}}}}
\newcommand*{\ttu}{\ensuremath{\mathtt{U}}}
\newcommand{\evenodd}{\mathtt{EvenOdd}}
\newcommand{\trios}{\mathtt{TRIOS}}
\newcommand{\up}{\mathtt{UP}}
\newcommand{\eq}{\mathtt{EQ}}
\newcommand{\expeq}{\mathtt{ExpEQ}}

\newcommand*{\bsection}[2]{\section[#1]{#1\label{#2}}\ignorespaces}

\ifdefined\endtheorem{}\else
  \newtheorem{theorem}{Theorem}[section]
\fi
\newcommand*{\btheorem}[1]{\begin{theorem}\label{#1}\ignorespaces}
\newcommand*{\etheorem}{\unskip\end{theorem}}

\ifdefined\endlemma{}\else
  \newtheorem{lemma}[theorem]{Lemma}
\fi
\newcommand*{\blemma}[1]{\begin{lemma}\label{#1}\ignorespaces}
\newcommand*{\elemma}{\unskip\end{lemma}}

\ifdefined\endcorollary{}\else
  \newtheorem{corollary}[theorem]{Corollary}
\fi
\newcommand*{\bcorollary}[1]{\begin{corollary}\label{#1}\ignorespaces}
\newcommand*{\ecorollary}{\unskip\end{corollary}}

\newcommand*{\bequations}{\begin{equation}\begin{array}{lcll}}
\newcommand*{\eequations}[1]{\end{array}\label{#1}\end{equation}\ignorespaces}

\newcommand*{\bclaim}{\begin{trivlist}\item[]\textbf{Claim.}\hspace{\labelsep}\em}
\newcommand*{\eclaim}{\unskip\end{trivlist}}

\newcommand*{\bitemize}{\begin{itemize}}
\newcommand*{\eitemize}{\unskip\end{itemize}}

\newcommand*{\bdescription}{\begin{description}}
\newcommand*{\edescription}{\unskip\end{description}}
\newcommand*{\ditem}[1]{\item[\mdseries{#1}]\ignorespaces}

\newcommand*{\bdisplay}{\[\begin{array}{lcll}}
\newcommand*{\edisplay}{\end{array}\]}

\ifdefined\endproof{}\else
  \newenvironment{proof}{\noindent\emph{Proof\/}:\hspace{\labelsep}\ignorespaces}{\bigbreak}
\fi
\newcommand*{\bproof}{\begin{proof}}
\newcommand*{\eproof}{\end{proof}}

\title{Classical Automata on Promise Problems%
  \thanks{A~preliminary version of this work was presented at the
    16\textsuperscript{th}~International Workshop on Descriptional
    Complexity of Formal Systems (DCFS~2014), August 5--8, 2014,
    Turku, Finland [vol.~8614 of LNCS, pp.~126--137, Springer-Verlag, 2014]\@.
    An~ArXiv version is at http://arxiv.org/abs/1405.6671.}}%
\author{Viliam Geffert\thanks{Partially supported by the Slovak grant contracts VEGA 1/0479/12
    and APVV-0035-10\@.} 
    \\
    Department of Computer Science,
    P.\,J.\,\v{S}af\'{a}rik University, Ko\v{s}ice, Slovakia 
    \\
   viliam.geffert@upjs.sk
   \\ \\
 Abuzer Yakary{\i}lmaz\thanks{Partially supported by CAPES with grant 88881.030338/2013-01,
    ERC Advanced Grant MQC, and FP7 FET project QALGO\@. Moreover,
    the part of the research work was done while Yakary{\i}lmaz was
    visiting Kazan Federal University\@.}%
    \\
    National Laboratory for Scientific Computing,
    Petr\'{o}polis, RJ, Brazil
    \\
    abuzer@lncc.br
}
    
\date{\today}%

\begin{document}

\maketitle

\begin{abstract}
Promise problems were mainly studied in quantum automata theory. Here we focus on state complexity of classical automata for promise problems. First, it was known that there is a family of unary promise problems solvable by quantum automata by using a single qubit, but the number of states required by corresponding one-way deterministic automata cannot be bounded by a constant. For this family, we show that even two-way nondeterminism does not help to save a single state. By comparing this with the corresponding state complexity of alternating machines, we then get a tight exponential gap between two-way nondeterministic and one-way alternating automata solving unary promise problems. Second, despite of the existing quadratic gap between Las Vegas realtime probabilistic automata and one-way deterministic automata for language recognition, we show that, by turning to promise problems, the tight gap becomes exponential. Last, we show that the situation is different for one-way probabilistic automata with two-sided bounded-error. We present a family of unary promise problems that is very easy for these machines; solvable with only two states, but the number of states in two-way alternating or any simpler automata is not limited by a constant. Moreover, we show that one-way bounded-error probabilistic automata can solve promise problems not solvable at all by any other classical model.
\end{abstract}

\bsection{Introduction}{s:intro}
Promise problem is a generalization of language recognition.
Instead of testing all strings over a given alphabet for language
membership, we focus only on a given subset of strings as
potential inputs that should be tested. The language under
consideration and its complement form this subset.

Promise problems have already provided several different
perspectives in computational complexity. (See the survey by
Goldreich~\cite{Gol06A}\@.) For example, it is not known whether
the class of languages recognizable by bounded-error probabilistic
polynomial-time algorithms has a complete problem, but the class
of promise problems solvable by the same type of algorithms has
some complete problems. A~similar phenomenon has been observed for
quantum complexity classes~\cite{Wat09A}\@. The first known result
on promise problems for restricted computational models we are
aware of was given by Condon and Lipton in 1989~\cite{CL89}\@.
They showed that a promise version of emptiness problem is
undecidable for one-way probabilistic finite automata. In the
literature, some separation results regarding restricted
computational models have also been given in the form of promise
problems. (See, e.g.,~\cite{DS90,DS92}\@.) The first result for
restricted quantum models was given by Murakami
\mbox{et\,al.\,}\cite{MNYW05}\@: There exists a promise problem
that can be solved by quantum pushdown automata exactly but not by
any deterministic pushdown automaton. Recently, Ambainis and
Yakary{\i}lmaz~\cite{AY12} showed that there is an infinite family
of promise problems which can be solved exactly just by tuning
transition amplitudes of a realtime two-state quantum finite
automaton (\qfa), whereas the size of the corresponding classical
automata grows above any fixed limit. Moreover, Rashid and
Yakary{\i}lmaz~\cite{RY14A} presented many superiority results of
\qfas\ over their classical counterparts. There are also some new
results on succinctness of
\qfas~\cite{ZQGLM13,GQZ14A,GQZ14B,ZGQ14A,BMP14B} and a new result
on quantum pushdown automata~\cite{Nak14A}\@.

\medbreak
In this paper, we turn our attention to classical models and
obtain some new results. In the next two sections, we give some
preliminaries and present some basic facts for the classical
models. Among others, we show that \iit{i}~the computational power
of deterministic, nondeterministic, alternating, and Las Vegas
probabilistic automata is the same, even after turning {}from the
classical language recognition to solving promise problems, and,
\iit{ii}~on promise problems, neither existing gaps of
succinctness for classical language recognition can be violated,
between any two of deterministic, nondeterministic, and
alternating automata models.

Then, in Section~\ref{s:expensive}, we consider a family of unary
promise problems given by Ambainis and Yakary{\i}l\-maz
in~\cite{AY12}, solvable by only two-state realtime quantum finite
automata. On the other hand, for solving this family by the use
classical automata, we show that the exact number of states for
one-way/two-way deterministic/nondeterministic automata is the
same. Thus, for this family, replacing the classical one-way
deterministic automata even by two-way nondeterministic automata
does not help to save a single state. However, for the same
family, already the one-way alternation does help, which gives us
the tight exponential gap between one-way alternating and two-way
sequential models of automata solving unary promise problems.

In Section~\ref{s:LV}, we show that the trade-off for the case of
Las Vegas probabilistic versus one-way deterministic automata is
different; by turning {}from language recognition to promise
problems the tight quadratic gap changes to a tight exponential
gap.

In Section~\ref{s:bounded-error}, we show how the situation
changes for finite state automata with two-sided bounded-error.
First, we present a probabilistically very easy family of unary
promise problems, i.e., solvable by one-way two-state automata
with bounded-error. Then, we show that this family is hard for
two-way alternating automata, and hence for any simpler classical
automata as well. Finally, we prove that, on promise problems,
bounded-error probabilistic automata are more powerful than any
other classical model.

\bsection{Preliminaries}{s:prel}
We assume the reader is familiar with the basic standard models of
finite state automata (see e.g.~\cite{HMU07})\@. For a more
detailed exposition related to probabilistic automata, the reader
is referred to~\cite{Pa71,Buk80}\@. Here we only recall some
models and introduce some elementary notation.

By~$\Sigma\str$, we denote the set of strings over an
alphabet~$\Sigma$\@. The set of strings of length~$n$ is denoted
by~$\Sigma^n$ and the unique string of length zero by~\eps\@.
By~$\nnumber$, we denote the set of all positive integers. The
cardinality of a finite set~$S$ is denoted by~$\card{S}$\@.

\medbreak
A~\emph{one-way nondeterministic finite automaton} with \eps-moves
(\owe\nfa, for short) is a quintuple
$\cla= (S,\Sigma,\lbr H,\lbr s\ini,S\acc)$, where $S$~is a finite
set of states, $\Sigma$~an input alphabet, $s\ini\in S$ an initial
state, and $S\acc\subseteq S$ a set of accepting (final) states.
Finally, $H\subseteq S\times(\Sigma\cup\set{\eps})\times S$ is a
set of \emph{transitions}, with the following meaning.
$s\trs{a} s'\in H$\@: if the next input symbol is~$a$, \cla~gets
{}from the state~$s$ to the state~$s'$ by reading~$a$ {}from the
input. $s\trs{\eps} s'\in H$\@: \cla~gets {}from~$s$ to~$s'$
without reading any symbol {}from the input.%
\footnote{Traditionally~\cite{HMU07}, the behavior is defined by a
  function $\delta: S\times(\Sigma\cup\set{\eps})\rightarrow 2^S$\@.
  However, a traditional transition $\delta(s,a)\ni s'$ can be
  easily converted to $s\trs{a} s'$ and vice versa. In the same
  way, all models introduced here are consistent with the
  traditional definitions. Using transitions instead of a
  $\delta$-function, we just make the notation more readable\@.}

The machine \cla~\emph{halts} in the state~$s$, if there are no
executable transitions in~$H$ at a time. Typically, this happens
after reading the entire input, but the computation may also be
blocked prematurely, by \emph{undefined transitions}, i.e., by
absence of transitions for the current state~$s$ and the next
input symbol~$a$ (or~$\eps$)\@.

An~\emph{accepting computation} starts in the initial
state~$s\ini$ and, after reading the entire input, it halts in any
accepting state $s'\in S\acc$\@. The language \emph{recognized}
(accepted) by~\cla, denoted by~$L(\cla)$, is the set of all input
strings having at least one accepting computation. The inputs with
no accepting computation are \emph{rejected}\@.

A~\emph{two-way nondeterministic finite automaton} (\tw\nfa) is
defined in the same way as \owe\nfa, but now \cla~can move in both
directions along the input. For these reasons, the transitions
in~$H$ are upgraded as follows. $s\trs{\!\!a\rightd} s'$,
$s\trs{\!\!a\leftd} s'$, $s\trs{\!\!a\stayd} s'$\@: if the current
input symbol is~$a$, \,\cla~gets {}from $s$ to~$s'$ and moves its
input head one position to the right, to the left, or keeps it
stationary, respectively. The input is enclosed in between two
special symbols \quo{$\vdash$} and~\quo{$\dashv$}, called the
\emph{left and right endmarkers}, respectively. By definition, the
input head cannot leave this area, i.e., there are no transitions
moving to the left of~$\vdash$ nor to the right of~$\dashv$\@.
\,\cla~starts in~$s\ini$ with the input head at the left endmarker
and accepts by halting in $s'\in S\acc$ anywhere along the input
tape.

\medbreak
A~\emph{deterministic finite automaton} (\owe\dfa\ or \tw\dfa) can
be obtained {}from \owe\nfa\ or \tw\nfa\ by claiming that the
transition set~$H$ does not allow executing more than one possible
transition at a time. Consequently, a \owe\dfa~\cla\ can have at
most one transition $s\trs{a} s'$, for each $s\in S$ and each
$a\in\Sigma$\@. In theory, one can consider \eps-transitions (even
though this feature is rarely utilized)\@: a~transition
$s\trs{\eps} s'$ implies that there are no other transitions
starting in the same state~$s$\@. Similar restrictions can be
formulated for \tw\dfas\@.

\medbreak
A~\emph{self-verifying finite automaton} (\owe\svfa\ or
\tw\svfa)~\cite{JP11} is a \owe\nfa\ or \tw\nfa~\cla\ which is
additionally equipped with $S\rej\subseteq S$, the set of
rejecting states disjoint {}from~$S\acc$, the set of accepting
states. The remaining states form
$S\neu= S\setminus(S\acc\cup S\rej)$, the set of neutral
\quo{don't know} states. In this model, \iit{i}~for each accepted
input, \cla~has at least one computation path halting in an
accepting state and, moreover, no path halts in a rejecting state,
\iit{ii}~for each rejected input, \cla~has at least one
computation path halting in a rejecting state and, moreover, no
path halts in an accepting state. In both these cases, some
computation paths may return \quo{don't know} answers, by halting
in neutral states or by getting into infinite loops.

\medbreak
A~\emph{one-way probabilistic finite automaton} with \eps-moves
(\owe\pfa) is defined in the same way as \owe\nfa, but the
transitions in~$H$ are upgraded with probabilities, as follows.
$s\trs{\!\!a,p} s'\in H$\@: if the current input symbol is~$a$,
\cla~gets {}from the state~$s$ to the state~$s'$ by reading this
symbol with probability $p\in [0,1]$\@. For $a=\eps$, no input
symbol is consumed. A~transition with $p=1$ may be displayed as
$s\trs{a} s'$, since the next step is deterministic; transitions
with $p=0$ can be completely erased, as not executable. By
definition,%
\footnote{Traditionally~\cite{Ra63}, the behavior of \pfas\ is
  defined by stochastic matrices\@.}
for each $s\in S$ and each $a\in\Sigma$, the sum of probabilities
over all transitions beginning in~$s$ and labeled by
$a'\in\set{a,\eps}$ must be equal to~$1$\@. The overall accepting
probability of~\cla\ on a given input $w\in\Sigma\str$, denoted
by~$f_{\cla}(w)$, is calculated over all accepting paths. Hence, the
overall rejecting probability is $1\!-\!f_{\cla}(w)$\@.

A~two-way version, \tw\pfa, was first introduced in~\cite{Ku73}\@.

A~\emph{Las Vegas probabilistic finite automaton} (Las Vegas
\owe\pfa\ or Las Vegas \tw\pfa)~\cite{HS01} is a \owe\pfa\ or
\tw\pfa\ of special kind, obtained {}from self-verifying automata
(i.e., {}from nondeterministic automata of special kind)\@. Thus,
transitions are upgraded with probabilities, and the state set is
partitioned into the sets of accepting, rejecting, and neutral
\quo{don't know} states. This gives three overall probabilities on
a given input; accepting, rejecting, and neutral.

\medbreak
A~\emph{one-way alternating finite automaton} with \eps-moves
(\owe\afa) is obtained {}from \owe\nfa\ by partitioning the state
set~$S$ into two disjoint subsets $S\exs$ and~$S\uns$, the sets of
\emph{existential} and \emph{universal} states, respectively. The
global rules are defined as is usual for alternating devices (see
e.g.~\cite{CKS81})\@: if, at the given moment, there are several
executable transitions, the machine~\cla\
\,\iit{i}~nondeterministically chooses one of them, if it is in an
existential state~$s$, but \iit{ii}~follows, in parallel, all
possible branches, if the current state~$s$ is universal.

For better readability, a nondeterministic switch {}from the
existential~$s$ into one of the states $s_1,\ldots,s_k$, by reading
an input symbol~$a$, will be displayed as
$s\trs{a} s_1\vee \cdots\vee s_k$, while a parallel branching {}from
the universal~$s$ to all these states as
$s\trs{a} s_1\wedge \cdots\wedge s_k$\@. The same applies for
$a=\eps$\@.

The input is accepted, if all branches in the nondeterministically
chosen computation subtree, rooted in the initial state at the
beginning of the input and embedded in the full tree of all possible
computation paths, halt in accepting states at the end of the
input.%
\footnote{\label{ft:boolean}The alternating automaton should not
  be confused with a \emph{Boolean finite automaton}, which is
  (quite inappropriately) also sometimes called \quo{alternating}
  in some literature. Instead of alternating existential and
  universal branching, the Boolean automaton can control
  acceptance by the use of arbitrarily complex Boolean functions.
  As an example, $s\trs{a} s_1\wedge(s_2\vee s_3)$ specifies that,
  for the current input symbol~$a$, the subtree rooted in the
  state~$s$ is successful if and only if the subtree rooted
  in~$s_1$ and at least one of those rooted in~$s_2,s_3$ are
  successful\@.}

Also this model was extended to a two-way version, \tw\afa\ (see
e.g.~\cite{Ge12})\@.

\medbreak
A~\emph{realtime automaton} (\rt\dfa, \rt\svfa, \rt\nfa, \rt\pfa,
\rt\afa,\dots\ ) is the corresponding one-way finite automaton%
\footnote{All realtime \emph{quantum} models use a classical input
  head and so they are not allowed to create a superposition of
  the head positions on the input tape\@.}
that executes at most a constant number of \eps-transitions in a
row. (After that, the next input symbol is consumed or the machine
halts\@.)

A~\emph{one-way \eps-free automaton} (\ow\dfa, \ow\svfa, \ow\nfa,
\ow\pfa, \ow\afa,\dots\ ) is the corresponding one-way finite
automaton with no \eps-transitions at all. (This is a special case
of a realtime machine, with the number of executed
\eps-transitions bounded by zero\@.)

It should be pointed out that, in \owe\nfas, the \eps-transitions
can be removed \emph{without increasing} the number of
states~\cite[Sect.~2.11]{Ku10}\@. Therefore, \ow\nfas, \rt\nfas,
and \owe\nfas\ agree in upper/lower bounds for states. The same
works for \ow\svfas, \rt\svfas, and \owe\svfas, as well as for
\ow\dfas, \rt\dfas, and~\owe\dfas\@. For this reason, we consider
only \ow\nfas, \ow\svfas, and \ow\dfas\@.

\medbreak
Occasionally, we shall also mention some restricted versions of
two-way machines, e.g., \emph{sweeping automata}, changing the
direction of the input head movement only at the
endmarkers~\cite{Sip80,KKM12}, or a very restricted version of
sweeping automaton called a \emph{restarting one-way automaton},
running one-way in an infinite loop~\cite{YS10B}\,---\,if the
computation does not halt at the right endmarker, then it jumps
back to the initial state at the beginning of the input.

\medbreak
A~\emph{promise problem} is a pair of languages
$\ttp= (\ttp\yes,\ttp\no)$, where
$\ttp\yes,\ttp\no\subseteq \Sigma\str$, for some~$\Sigma$, and
$\ttp\yes\cap\ttp\no= \emptyset$~\cite{Wat09A}\@.

The promise problem~\ttp\ is \emph{solved} by a finite
automaton~\clp, if \clp\ accepts each $w\in\ttp\yes$ and rejects
each $w\in\ttp\no$\@. (That is, we do not have to worry about the
outcome on inputs belonging neither to~$\ttp\yes$ nor
to~$\ttp\no$\@.) If $\ttp\yes\cup\ttp\no= \Sigma\str$ and \ttp~is
solved by~\clp, we have a classical language recognition, and say
that $\ttp\yes$~is \emph{recognized} by~\clp\@.

If \clp~is a probabilistic finite automaton, we say that \clp\
solves~\ttp\ with \emph{error bound} $\eps\in [0,\frac{1}{2})$, if
\clp\ accepts each $w\in\ttp\yes$ with probability at least
$1\!-\!\eps$ and rejects each $w\in\ttp\no$ with probability at
least $1\!-\!\eps$\@. (The remaining probability goes to erroneous
answers\@.) \clp~solves \ttp\ with a \emph{bounded-error}, if, for
some~\eps, it solves~\ttp\ with the error bound~\eps\@. If
$\eps=0$, the problem is solved \emph{exactly}\@.

A~special case of bounded-error is a \emph{one-sided bounded-error},
where either each $w\in\ttp\yes$ is accepted with probability~$1$ or
each $w\in\ttp\no$ is rejected with probability~$1$\@.

If \clp\ is a Las Vegas probabilistic finite automaton, we say
that \clp\ solves~\ttp\ with a \emph{success probability}
$p\in (0,1]$, if \iit{i}~for each $w\in\ttp\yes$,
\,\clp~accepts~$w$ with probability at least~$p$ and never rejects
(i.e., the probability of rejection is~$0$) and \iit{ii}~for each
$w\in\ttp\no$, \,\clp~rejects~$w$ with probability at least~$p$
and never accepts. (In both these cases, the remaining probability
goes to \quo{don't know} answers\@.)

\bsection{Basic Facts on Classical Automata}{s:basics}
We continue with some basic facts regarding classical automata. We start with a basic observation: The cardinality of the class of the promise problems defined by \ow\dfas~ is uncountable. We can show this fact even for the class defined by a fixed 2-state \ow\dfa~ on unary languages. Let $ \mathtt{L} \subseteq \mathcal{N} $ be any language defined on natural numbers. Then, the following unary promise problem $ \mathtt{P(L)} $ can be solvable by a 2-state \ow\dfa, say $ \mathcal{D} $:
\[
	\begin{array}{llll}
		\mathtt{P_{yes}(L)} & = & \{ a^{2n}  & \mid n \in \mathtt{L} \}
		\\
		\mathtt{P_{no}(L)} & = & \{ a^{2n+1} & \mid n \in \mathtt{L} \}
	\end{array} ,
\]
where $ \mathcal{D} $ has two states, the initial and only accepting state $s_1$  and $ s_2 $, and, the automaton alternates between $s_1$ and $s_2$ for each input symbol. There is a one-to-one correspondence between $ \{ \mathtt{L} \mid \mathtt{L} \subseteq \mathcal{N}  \} $ and $ \{ \mathtt{P(L)} \mid \mathtt{L} \subseteq \mathcal{N} \} $. Therefore, the cardinality of the class of the promise problems solvable by $ \mathcal{D} $ is uncountable. Moreover, if $ \mathtt{L} $ is an uncomputable language, then both $ \mathtt{P_{yes}(L)} $ and $ \mathtt{P_{no}(L)} $ are uncomputable. The interested reader can easily find some other promise problems (solvable by very small \ow\dfas) whose yes- and no-instances satisfy some special properties (i.e. non-context-free but context-sensitive, NP-complete, semi-decidable, etc.). One trivial construction for a binary language $ \mathtt{L} \in \{ a,b \}^* $ is to define the promise problem as $ \mathtt{P(L)} = ( \{ aw \mid w \in \mathtt{L} \},\{ bw \mid w \in \mathtt{L} \} ) $. (A promise problem solvable by \ow\dfas~ whose yes- and no-instances are non-regular languages previously given in \cite{GQZ14B}.) Despite of the above facts, all trade-offs established between various models of automata stay valid also for promise problems.

First, we show that the classes of promise problems solvable by
deterministic, nondeterministic, alternating, and Las Vegas
probabilistic finite automata are identical.

\btheorem{t:afa-to-dfa}
If a promise problem $\ttp=(\ttp\yes,\ttp\no)$ can be solved by a
\tw\afa~\cla\ with $n$ states, it can also be solved by a
\ow\dfa~\cld\ with $t(n)\le 2^{n\cdot 2^n}$ states.
\etheorem
\bproof
Let~\ttr\ denote the regular language recognized by~\cla\@. (That
is, even though we are given a \tw\afa~\cla\ for solving a promise
problem~\ttp, such machine is still associated with some classical
regular language $L(\cla)=\ttr$\@.) The complement of this
language will be denoted by~\ttrc\@. Since the problem \ttp\ is
solvable by~\cla, we have that $\ttp\yes\subseteq \ttr$ and
$\ttp\no\subseteq \ttrc$\@.

Now, by~\cite{LLS84}, each \tw\afa~\cla\ with $n$ states can be
transformed into an equivalent \ow\dfa~\cld\ with
$t(n)\le 2^{n\cdot 2^n}\!$ states. This gives that
$\ttp\yes\subseteq \ttr= L(\cld)$ and
$\ttp\no\subseteq \ttrc= \overline{L(\cld)}$, that is,
\cld~accepts each $w\in\ttp\yes$ and rejects each $w\in\ttp\no$\@.
Therefore, \cld~can be used for solving~\ttp\@.
\eproof

The proof of the above theorem can be easily updated for other
classical models of automata\,---\,using the corresponding
trade-off~$t(n)$ for the conversion known {}from the
literature\,---\,by which we obtain the following corollary.

\bcorollary{c:not-violated}
Any trade-off~$t(n)$ in the number of states for language
recognition, between any two of deterministic, nondeterministic,
or alternating automata models, is also a valid trade-off for
promise problems.
\ecorollary

Next, we show that neither Las~Vegas \pfas\ gain any computational
power.

\btheorem{t:lv-to-dfa}
If a promise problem $\ttp=(\ttp\yes,\ttp\no)$ can be solved by a
Las Vegas \tw\pfa~\clp\ with $n$ states and any success
probability $p>0$, it can also be solved by a \ow\dfa~\cld\ with
$t(n)\le 2^{n^2+n}$ states.
\etheorem
\bproof
First, for each $w\in\ttp\yes$, the machine~\clp\ has at least one
accepting computation path and no rejecting path. Conversely, for
each $w\in\ttp\no$, the machine has at least one rejecting path
and no accepting path. In both cases, the paths that are not
successful return a \quo{don't know} answer.

Therefore, by removing the probabilities and converting neutral
\quo{don't know} states into rejecting states, we obtain an
$n$-state \tw\nfa~\cln\ recognizing a regular language~\ttr\ such
that $\ttp\yes\subseteq \ttr$ and
$\ttp\no\cap\ttr\ = \emptyset$\@. Therefore,
$\ttp\no\subseteq \ttrc$\@.

Now, by~\cite{Ka05m}, the \tw\nfa~\cln\ can be converted to an
equivalent \ow\dfa~\cld\ with the number of states bounded by
$t(n)\le
  \sum_{i=0}^{n-1} \sum_{j=0}^{n-1} ({n\atop i})({n\atop j})(2^i\!-\!1)^j\le
  2^{n^2+n}$,
which gives $\ttp\yes\subseteq \ttr= L(\cld)$ and
$\ttp\no\subseteq \ttrc= \overline{L(\cld)}$\@. That is,
\cld~accepts each $w\in\ttp\yes$ and rejects each $w\in\ttp\no$,
and hence it can be used for solving~\ttp\@.
\eproof

A~similar idea can be used for Las Vegas \owe\pfas\@:

\bcorollary{c:lv-to-dfa}
If a promise problem $\ttp=(\ttp\yes,\ttp\no)$ can be solved by a
Las Vegas \owe\pfa~\clp\ with $n$ states and any success
probability $p>0$, it can also be solved by a \ow\dfa~\cld\ with
$t(n)\le 1\!+\!3^{(n-1)/3}\le 2^{0.529\cdot n}$ states.
\ecorollary
\bproof
By removing the probabilities, we first obtain an $n$-state
\owe\svfa~\cln\ recognizing a regular language~\ttr\ such that
$\ttp\yes\subseteq \ttr$ and $\ttp\no\subseteq \ttrc$\@. Then,
after elimination of \eps-transitions by the method described
in~\cite[Sect.~2.11]{Ku10} (still not increasing the number of
states), we have a \ow\svfa\ that can be converted to an
equivalent \ow\dfa~\cld\ with $t(n)\le 1+3^{(n-1)/3}$
states~\cite{JP11}\@. The resulting \ow\dfa\ is then used for
solving~\ttp\@.
\eproof

In Section~\ref{s:LV}, we show that the corresponding lower bound
is also exponential.

\bsection{A~Classically Expensive Unary Promise Problem}{s:expensive}
Recently, Ambainis and Yakary{\i}lmaz~\cite{AY12} presented a
family of promise problems $\evenodd(k)$, for
\mbox{$k\!\in\!\nnumber$}, where
\bdisplay
  \evenodd\yes(k) &=& \set{ a^{m\cdot 2^{k}} \st m \mbox{ is even} } \,,\\
  \evenodd\no(k)  &=& \set{ a^{m\cdot 2^{k}} \st m \mbox{ is odd} } \,,
\edisplay
such that, for each~$k$, the promise problem $\evenodd(k)$ can be
solved exactly by only a two-state realtime quantum finite
automaton. On the other hand, for \ow\dfas, $2^{k+1}$~states are
necessary and also sufficient. Later, it was shown that
$2^{k+1}$~states are also necessary in any bounded-error
\ow\pfa~\cite{RY14A} and that \tw\dfas\ must use at least
$2^{k+1}\!-\!1$ states~\cite{BMP14B}\@.%
\footnote{Recently~\cite{AGKY14A}, it was also shown that the
  width of any nondeterministic or bounded-error stable \obbd\
  cannot be smaller than~$2^{k+1}$, for a functional version of
  $\evenodd(k)$\@.}

Here we show that even two-way nondeterminism does not help us to
save a single state for this promise problem; and hence
$2^{k+1}$~states is the \emph{exact} value for one-way/two-way
deterministic/nondeterministic automata. After that, we show that
alternation does help to reduce the number of states
to~$\Theta(k)$, already by the use of one-way \mbox{\eps-}free
machines. More precisely, $\evenodd(k)$ can be solved by a
\ow\afa\ with $11k\!-\!14$ states (for $k\ge 3$); the corresponding
lower bound for \ow\afas\ is~$k\!+\!1$\@. For two-way
machines,~\tw\afas, the exact lower bound is an open problem, but
we know that it must grow in~$k$, which is an easy consequence of
Theorems \ref{t:afa-to-dfa} and~\ref{t:nfa-even-odd}\@. Similarly,
we do not know whether bounded-error \tw\pfas\ can work with fewer
than $2^{k+1}$ states.

\medbreak
The method used for proving the lower bound in the following
theorem is not new\,---\,inspired
by~\cite{SHL65,CHR91,Ge91}\,---\,but it needs some modifications
on promise problems.

\btheorem{t:nfa-even-odd}
For each $k\in\nnumber$, any \tw\nfa\ for solving the promise problem
$\evenodd(k)$ needs at least $2^{k+1}$ states.
\etheorem
\bproof
Suppose, for contradiction, that the promise problem can be solved
by a \tw\nfa\ \cln\ with $\card{S}<2^{k+1}$ states.

Consider now the input~$a^{2^{k+1}}\!$\@. Clearly, it must be
accepted by~\cln, and hence there must exist at least one
accepting computation path, so we can fix one such path. (Besides,
being nondeterministic, \cln~can have also other paths, not all
of them necessarily accepting, but we do not care for them\@.)
Along this fixed path, take the sequence of states
\bdisplay
  & q_0,\,p_1,q_1,\,p_2,q_2,\,p_3,q_3,\ldots,p_r,q_r,\,p_{r+1} \,,
\edisplay
where $q_0$~is the initial state, $p_{r+1}$~is an accepting state,
and all the other states $\set{p_i\cup q_i\st 1\le i\le r}$ are at
the left/right endmarkers of the input~$2^{k+1}$, such that:
\bitemize
  \item If $p_i$~is at the left (resp., right) endmarker, then
    $q_i$ is at the opposite endmarker, and the path {}from $p_i$
    to~$q_i$ traverses the entire input {}from left to right
    (resp., right to left), not visiting any of endmarkers in the
    meantime.
  \item The path between $q_i$ and~$p_{i+1}$ starts and ends at
    the same endmarker, possibly visiting this endmarker several
    times, but not visiting the opposite endmarker. Such a path is
    called a \quo{U-turn}. This covers also the case of
    $q_i=p_{i+1}$ with zero number of executed steps in between.
\eitemize

Now, the path connecting $p_i$ with~$q_i$ must visit all input
tape positions in the middle of the input~$a^{2^{k+1}}\!$, and
hence \cln\ must enter $2^{k+1}$ states, at least one state per
each input tape position. However, since $\card{S}< 2^{k+1}$,
some state $r_i$ must be repeated. That is, between $p_i$
and~$q_i$, there must exist a loop, starting and ending in the
same state~$r_i$, traveling some $\ell_i$~positions to the right,
not visiting any of the endmarkers. For each traversal, we fix one
such loop (even though the path {}from $p_i$ to~$q_i$ may contain
many such loops)\@. Note that the argument for traveling across
the input {}from right to left is symmetrical. Since this loop is
in the middle of the traversal, we have that $\ell_i<2^{k+1}$\@.
But then $\ell_i$ can be expressed in the form
\bdisplay
  \ell_i &=& \gamma_i\!\cdot\!2^{\alpha_i} ,
\edisplay
where $\gamma_i>0$ is odd (not excluding $\gamma_i=1$) and
$\alpha_i\in \set{0,1,\ldots,k}$\@. Note that if
$\alpha_i\ge k\!+\!1$, we would have a contradiction with
$\ell_i<2^{k+1}$\@.

Now, consider the value
\bdisplay
  \ell &=& \lcm\set{ \ell_1,\ell_2,\ldots,\ell_r } \,,
\edisplay
the least common multiple of all fixed loops, for all traversals
of the input~$a^{2^{k+1}}\!$\@. Clearly, $\ell$~can also be
expressed in the form
\bdisplay
  \ell &=& \gamma\!\cdot\!2^{\alpha} ,
\edisplay
where $\gamma>0$ is odd and $\alpha\in \set{0,1,\ldots,k}$
\,(actually, $\gamma= \lcm\set{\gamma_1,\ldots,\gamma_r}$ and
$\alpha= \max\set{\alpha_1,\ldots,\alpha_r}$)\@.

We shall now show that the machine \cln\ must also accept the
input
\bdisplay
  & a^{2^{k+1}+\ell\cdot 2^{k-\alpha}} .
\edisplay
First, if $q_i$ and~$p_{i+1}$ are connected by a U-turn path on
the input~$a^{2^{k+1}}\!$, they will be connected also on the
input $a^{2^{k+1}+\ell\cdot 2^{k-\alpha}}$ since such path does
not visit the opposite endmarker and
$2^{k+1}+\ell\!\cdot\!2^{k-\alpha}\ge 2^{k+1}$\@. Second, if $p_i$
and~$q_i$ are connected by a left-to-right traversal
along~$a^{2^{k+1}}\!$, they will stay connected also along the
input $a^{2^{k+1}+\ell\cdot 2^{k-\alpha}}\!$\@. This only requires
to repeat the loop of the length~$\ell_i$ beginning and ending in
the state~$r_i$\@. Namely, we can make
$\frac{\ell}{\ell_i}\cdot 2^{k-\alpha}$ more iterations. Note that
$\ell= \lcm\set{\ell_1,\ldots,\ell_r}$ and hence $\ell$ is
divisible by~$\ell_i$, for each $i=1,\ldots,r$, and that
$\alpha\le k$, which gives that
$\frac{\ell}{\ell_i}\cdot 2^{k-\alpha}$ is an integer. Note also
that these $\frac{\ell}{\ell_i}\cdot 2^{k-\alpha}$ additional
iterations of the loop of the length~$\ell_i$ travel exactly
$(\frac{\ell}{\ell_i}\cdot 2^{k-\alpha})\times\ell_i =
  \ell\!\cdot\!2^{k-\alpha}$
additional positions to the right. Thus, if \cln~has an
accepting path for~$a^{2^{k+1}}\!$, it must also have an accepting
path for $a^{2^{k+1}+\ell\cdot 2^{k-\alpha}}$ (just by a
straightforward induction on~$r$)\@. Therefore, \cln~accepts
$a^{2^{k+1}+\ell\cdot 2^{k-\alpha}}\!$\@. (Actually, \cln~can
have many other paths for this longer input, but they cannot rule
out the accepting decision of the path constructed above\@.)

However, the input $a^{2^{k+1}+\ell\cdot 2^{k-\alpha}}$ should be
rejected, since
$2^{k+1}+ \ell\!\cdot\!2^{k-\alpha}=
  2^{k+1}+ \gamma\!\cdot\!2^{\alpha}\!\cdot\!2^{k-\alpha}=
  (2\!+\!\gamma)\cdot 2^k$,
where $\gamma$ is odd. This is a contradiction. So, \cln~must
have at least $2^{k+1}$ states.
\eproof

The above argument cannot be extended to alternating automata:
using \quo{cooperation} of several computation paths running in
parallel, the number of states can be reduced {}from $2^{k+1}$
to~$O(k)$, even by the use of one-way \eps-free machines. We first
present a conceptually simpler realtime version.

\blemma{l:afa-even-odd}
For each $k\in\nnumber$, the language $\evenodd\yes(k)$ can be
recognized\,---\,and hence the promise problem $\evenodd(k)$ can
be solved\,---\,by an \rt\afa~\cla\ with $7k\!+\!2$ states.
\elemma
\bproof
Recall that
$\evenodd\yes(k)= \set{a^{m\cdot 2^{k}} \st m \mbox{ is even}}$,
and hence the unary input string~$a^n$ is in this language if and
only if $n$~is an integer multiple of~$2^{k+1}$\@. In turn, this
holds if and only if $b_{n,k}= b_{n,k-1}= \cdots= b_{n,0}= 0$,
where $b_{n,i}$~denotes the $i$\xth~bit in the binary
representation of the number~$n$\@. (The bit positions are
enumerated {}from right to left, beginning with zero for the least
significant bit\@.) That is, written in binary, $n$~should be of
the form~$\gamma 0^{k+1}$, for some $\gamma\in\set{0,1}\str$\@.

Based on this, the \rt\afa~\cla\ counts the length of the input
modulo~$2^{k+1}$ and verifies, for the binary representation of
this length, that the last $k\!+\!1$ bits are all equal to zero.
The important detail is that \cla~counts \emph{down}, starting
{}from zero. That is, along the input, the value of the counter
changes as follows:
\bdisplay
  \ell &:=& 0,\,2^{k+1}\!-\!1,\,2^{k+1}\!-\!2,\,\ldots,
    \,2,\,1,\,0,\,2^{k+1}\!-\!1,\,\ldots
\edisplay
Thus, the counter $\ell$ is interpreted as the number of input
symbols not processed yet (and taken modulo~$2^{k+1}$)\@. The
machine~\cla\ accepts if and only if this value is zero when it
halts at the end of the input.

Implemented deterministically, this counting requires $2^{k+1}$
states. However, an alternating automaton can use several
processes running in parallel. (See~\cite{RY14B} for a similar
idea\@.) Each parallel process will keep only \emph{one bit} of
the current value of~$\ell$\@. That is, for
$i\in\set{k,k\!-\!1,\ldots,0}$, we shall use the states
\quo{$i_{0}$} and~\quo{$i_{1}$} to indicate that $i$\xth\ bit in
the current value of~$\ell$ is cleared to zero or set to one,
respectively. This will hold for each \quo{proper} sequence of
existential guesses along each input. The correctness of this
guessing will be verified when \cla\ halts at the end of the
input. (Actually, the alternating tree of computation paths will
be highly redundant; the same bit will be remembered by a huge
number of activated identical copies of $i_{0}$ or~$i_{1}$,
running in parallel\@.) Clearly, for each symbol~$a$ along the
input, we need to decrement the counter~$\ell$ by one and hence to
update properly the corresponding bits. This is implemented by the
use of some additional auxiliary states.

We are now ready to present the set of states for the
\rt\afa~\cla, together with their brief description:
\bdescription
  \ditem{$i_{x}$,} for $i=k,\ldots,0$ and $x\in\set{0,1}$\@: the
    $i$\xth\ bit of~$\ell$ is~$x$
    \ ($b_{\ell,i}=x$),
  \ditem{$i_{x,\allone}$,} for $i=k,\ldots,1$ and $x\in\set{0,1}$\@:
    the $i$\xth\ bit of~$\ell$ is~$x$ and, moreover, all bits to
    the right are set to~$1$
    \ ($b_{\ell,i}=x$ and $b_{\ell,j}=1$ for each $j<i$),
  \ditem{$i_{x,\exzero}$,} for $i=k,\ldots,1$ and $x\in\set{0,1}$\@:
    the $i$\xth\ bit of~$\ell$ is~$x$ and, moreover, at least one
    bit to the right is set to~$0$
    \ ($b_{\ell,i}=x$ and $b_{\ell,j}=0$ for some $j<i$),
  \ditem{$i_{\exzero}$,} for $i=k,\ldots,2$\@: the $i$\xth\ bit
    of~$\ell$ may be quite arbitrary, but at least one bit to the
    right is set to~$0$
    \ ($b_{\ell,j}=0$ for some $j<i$),
  \ditem{$s\ini$\@:} the initial state.
\edescription

All states of type $i_{0},\,i_{1},\,i_{\exzero}$ are existential,
whereas those of type
$i_{0,\allone},\lbr \,i_{1,\allone},\lbr \,i_{0,\exzero},\lbr
  \,i_{1,\exzero},\lbr \,s\ini$
are universal. The states of type $i_{0},\,s\ini$ are accepting
and none of the remaining states is accepting.

Clearly, the total number of states is
$2(k\!+\!1) +2k +2k +(k\!-\!1) +1= 7k\!+\!2$\@.

Transitions in~\cla\ are defined as follows:
\bdescription
  \ditem{$i_{x}\trs{a} i_{x,\exzero}\vee i_{1-x,\allone}$,} for
    $i=k,\ldots,1$ and $x\in\set{0,1}$\@: The $i$\xth\ bit
    of~$\ell$ is set to~$x$ if and only if either \iit{i}~the
    $i$\xth\ bit of $\ell'=\ell\!-\!1$ is set to the same
    value~$x$, provided that at least one bit to the right
    in~$\ell'$ is set to zero, that is, $b_{\ell',j}=0$ for some
    $j<i$, or \iit{ii}~the $i$\xth\ bit of~$\ell'$ is flipped to
    the complementary value~$1\!-\!x$, provided that all bits the
    right in~$\ell'$ are set to one, that is, $b_{\ell',j}=1$ for
    each $j<i$\@. Therefore, in the state~$i_{x}$, branching
    existentially while reading the next input symbol~$a$,
    \cla~guesses between switching to $i_{x,\exzero}$ or
    to~$i_{1-x,\allone}$\@.
  \ditem{$0_{x}\trs{a} 0_{1-x}$,} for $x\in\set{0,1}$, a special
    case of~$i_{x}$ for $i=0$\@: Transitions for the rightmost bit
    are simplified, since there is no bit position with $j<i$ and
    hence we do not need to use
    $0_{0,\allone},\,0_{1,\allone},\lbr \,0_{0,\exzero},\,0_{1,\exzero}$\@.
    Therefore, in~$0_{x}$, the machine reads the next input
    symbol~$a$ deterministically, by switching to~$0_{1-x}$\@.
  \ditem{$i_{x,\allone}\trs{\eps} i_{x}\wedge (i\!-\!1)_{1}\wedge
      \cdots\wedge 0_{1}$,}
    for $i=k,\ldots,1$ and $x\in\set{0,1}$\@: In the auxiliary
    state~$i_{x,\allone}$, the machine~\cla\ has to verify the bit
    setting at the corresponding positions in~$\ell'$, namely,
    that $b_{\ell',i}=x$ and that $b_{\ell',j}=1$ for each
    $j<i$\@. Thus, branching universally with \eps-transitions for
    each bit position $j\le i$, it splits into parallel processes
    $i_{x},(i\!-\!1)_{1},\ldots,0_{1}$\@.
  \ditem{$i_{x,\exzero}\trs{\eps} i_{x}\wedge i_{\exzero}$,} for
    $i=k,\ldots,2$ and $x\in\set{0,1}$\@: The role of the
    auxiliary state~$i_{x,\exzero}$ is similar to that
    of~$i_{x,\allone}$, verifying the bit setting in~$\ell'$\@.
    This time we verify that \iit{i}~$b_{\ell',i}=x$ and that
    \iit{ii}~$b_{\ell',j}=0$ for at least one $j<i$\@. In order
    not to mix universal and existential decisions, we delay the
    condition~\iit{ii} until the next computation step, and hence
    \cla~branches universally in~$i_{x,\exzero}$, splitting into
    two parallel processes $i_{x}$ and~$i_{\exzero}$, by the use
    of \eps-transitions.
  \ditem{$1_{x,\exzero}\trs{\eps} 1_{x}\wedge 0_{0}$,} for
    $x\in\set{0,1}$, a special case of~$i_{x,\exzero}$ for
    $i=1$\@: Transitions are simplified, since we can work
    without~$1_{\exzero}$\@. Therefore, in~$1_{x,\exzero}$, the
    machine splits universally into $1_{x}$ and~$0_{0}$, using
    \eps-transitions.
  \ditem{$i_{\exzero}\trs{\eps} (i\!-\!1)_{0}\vee \cdots\vee \,0_{0}$,}
    for $i=k,\ldots,2$\@: In the auxiliary state~$i_{\exzero}$, we
    verify that $b_{\ell',j}=0$ for at least one $j<i$\@.
    Therefore, branching existentially with \eps-transitions,
    \cla~guesses which bit is set to zero, switching to~$j_{0}$,
    for some $j<i$\@.
  \ditem{$s\ini\trs{a} k_{1,\allone}$\@:} At the very beginning,
    $\ell=n$\@. Therefore, for each accepted input~$a^n$, the
    initial value of~$\ell$ must be an integer multiple
    of~$2^{k+1}$, with the last $k\!+\!1$ bits all equal to zero.
    Thus, after reading the first input symbol~$a$, we should get
    $\ell'=\ell\!-\!1$ in which all these bits are set to one. For
    this reason, in the initial state, the machine reads the first
    input symbol~$a$ by switching deterministically
    to~$k_{1,\allone}$\@.
\edescription

It is easy to see {}from the above transitions that \cla\ is a
realtime machine: After reading an input symbol, a computation
path can pass through at most two \eps-transitions in a row, after
which it gets to a state~$i_{x}$, for some $i=k,\ldots,0$ and some
$x\in\set{0,1}$\@. In~$i_{x}$, the machine waits for a next input
symbol.

To show that \cla\ indeed recognizes $\evenodd\yes(k)$, we need to
prove the following claim.

\bclaim
For each $\ell\ge 0$, each $i=k,\ldots,0$, and each
$x\in\set{0,1}$, the \rt\afa~\cla\ has an accepting alternating
subtree of computation paths rooted in the state~$i_{x}$ at the
input position~$\ell$ (measured {}from the end of the input) if and
only if the $i$\xth~bit of~$\ell$ is equal to~$x$\@.
\eclaim
The claim is proved by induction on $\ell=0,1,2,\ldots$

First, let $\ell=0$, that is, \cla~has already processed all input
symbols. Clearly, all bits of~$\ell$ are equal to zero, i.e.,
$b_{\ell,i}=0$ for each~$i$\@. Now, since there are no more input
symbols to be processed and there are no \eps-transitions starting
in the state~$i_{x}$, the machine \cla\ must halt in~$i_{x}$\@.
Therefore, \cla~has an accepting alternating subtree rooted
in~$i_{x}$ at the end of the input if and only if $i_{x}$~is an
accepting state. By definition of accepting states in~\cla\ (the
states of type~$i_{0}$ are accepting but those of type~$i_{1}$ are
not), this holds if and only if $x=0$, which in turn holds if and
only if $x=0=b_{\ell,i}$, the $i$\xth~bit of~$\ell$\@.

Second, by induction, assume that the claim holds for
$\ell'=\ell\!-\!1$\@. Now, by definition of the transitions in the
state~$i_{x}$ for $i>0$, taking into account subsequent
\mbox{\eps-}transitions passing through auxiliary states, it can
be seen%
\footnote{With obvious modifications, the argument presented here
  can be easily extended also for the case of $i=0$ (the rightmost
  bit), which we leave to the reader\@.}
that \cla\ has an accepting alternating subtree of computation
paths rooted in the state~$i_{x}$ at the input position~$\ell$ if
and only if either \iit{i}~\cla~has an accepting alternating
subtree rooted in the state~$i_{x}$ at the input position
$\ell'=\ell\!-\!1$ and, moreover, the same holds for at least one
subtree rooted in a state~$j_{0}$ at the position~$\ell'$, for
some $j<i$, or \iit{ii}~\cla~has an accepting alternating subtree
rooted in~$i_{1-x}$ at the input position~$\ell'$ and, moreover,
the same holds for all subtrees rooted in~$j_{1}$ at the
position~$\ell'$, for each $j<i$\@. However, by induction, using
$\ell'=\ell\!-\!1$, this statement holds if and only if either
\iit{i}~$b_{\ell',i}=x$ and, moreover, $b_{\ell',j}=0$, for some
$j<i$, or \iit{ii}~$b_{\ell',i}=1\!-\!x$ and, moreover,
$b_{\ell',j}=1$, for each $j<i$\@. Using $\ell=\ell'\!+\!1$, it is
easy to see that this holds if and only if $b_{\ell,i}=x$\@.

This completes the proof of the claim.

\medbreak
Now, recall that the machine~\cla\ reads the first input
symbol~$a$ by switching deterministically {}from the initial state
to the state~$k_{1,\allone}$\@. After that, branching universally
with \eps-transitions for each bit position $j\le k$, the
computation splits into parallel processes
$k_{1},(k\!-\!1)_{1},\ldots,0_{1}$\@. Thus, using the above claim,
\cla~accepts~$a^n$ with $n>0$ if and only if, in $\ell=n\!-\!1$,
the last $k\!+\!1$ bits are all equal to~$1$\@. Clearly, this
holds if and only if the last $k\!+\!1$ bits in~$n$ are all equal
to~$0$, that is, if and only if $n$~is and integer multiple
of~$2^{k+1}$\@. This gives that the language recognized by~\cla\
agrees with $\evenodd\yes(k)$ on all inputs of length $n>0$\@.

For the special case of $n=0$, we have that
$a^0= a^{0\cdot 2^{k}}\in \evenodd\yes(k)$, since $m=0$ is even.
However, $a^0=\eps$ is accepted by~\cla\ due to the fact that the
initial state~$s\ini$ is also an accepting state, which completes
the argument.
\eproof

By increasing slightly the number of states, we can replace the
realtime alternating machine {}from the above lemma by a one-way
\afa\ not using any \mbox{\eps-}transitions:

\btheorem{t:afa-even-odd}
For each $k\ge 3$, the language $\evenodd\yes(k)$ can be
recognized\,---\,and hence the promise problem $\evenodd(k)$ can
be solved\,---\,by a \ow\afa~$\cla''$ with $11k\!-\!14$ states.
\etheorem
\bproof
By inspection of the \rt\afa~\cla\ constructed in
Lemma~\ref{l:afa-even-odd}, it can be seen that \cla\ can read an
input symbol only by transitions starting in the states of type
$i_{x},\,s\ini$\@. After reading, the computation path can pass
through at most two subsequent \eps-transitions until it gets to a
state of type~$i_x$, ready to read {}from the input again.

First, as an intermediate product, we replace~\cla\ by an
equivalent machine~$\cla'$ in which reading an input symbol is
always followed by \emph{exactly three} subsequent
\mbox{\eps-}transitions, along each computation path. This only
requires to delay artificially the moment when the automaton gets
to a state of type~$i_x$ by the use of some additional states,
namely, the following additional states:
\bdescription
  \ditem{$i'_{x},i''_{x}$,} for $i=k,\ldots,0$ and $x\in\set{0,1}$\@:
    delay switching to the state~$i_{x}$ by one or two steps,
    respectively\,---\,implemented by transitions
    $i''_{x}\trs{\eps} i'_{x}\trs{\eps} i_{x}$,
  \ditem{$0'''_{x}$,} for $x\in\set{0,1}$\@: delay switching to
    the state~$0_{x}$ by three steps\,---\,implemented by
    $0'''_{x}\trs{\eps} 0''_{x}$\@.
\edescription
Taking also into account the states used by the original
machine~\cla, the total number of states increases to
$(7k\!+\!2) +4(k\!+\!1) +2= 11k\!+\!8$\@.

Transitions for the \quo{original} states (cf.~construction in
Lemma~\ref{l:afa-even-odd}) are modified as follows:
\bitemize
  \item[] $i_{x}\trs{a} i_{x,\exzero}\vee i_{1-x,\allone}$, for
    $i=k,\ldots,1$ and $x\in\set{0,1}$ \ (not modified),
  \item[] $0_{x}\trs{a} 0'''_{1-x}$, for $x\in\set{0,1}$,
  \item[] $i_{x,\allone}\trs{\eps} i''_{x}\wedge
      (i\!-\!1)''_{1}\wedge \cdots\wedge 0''_{1}$,
    for $i=k,\ldots,1$ and $x\in\set{0,1}$,
  \item[] $i_{x,\exzero}\trs{\eps} i''_{x}\wedge i_{\exzero}$, for
    $i=k,\ldots,2$ and $x\in\set{0,1}$,
  \item[] $1_{x,\exzero}\trs{\eps} 1''_{x}\wedge 0''_{0}$, for
    $x\in\set{0,1}$,
 \item[] $i_{\exzero}\trs{\eps} (i\!-\!1)'_{0}\vee \cdots\vee \,0'_{0}$,
    for $i=k,\ldots,2$,
  \item[] $s\ini\trs{a} k_{1,\allone}$ \ (not modified)\@.
\eitemize
Since $\cla'$ and~\cla\ share the same set of accepting states, it
should be obvious that they recognize the same language.

Second, replace all \eps-transitions in~$\cla'$ by transitions
reading the input symbol~$a$\@. That is, each transition of type
$s_1\trs{\eps} s_2$ is replaced by $s_1\trs{a} s_2$; the
transitions already reading input symbols do not change; nor does
the set of accepting states. This way we obtain a one-way
\afa~$\cla''$ without any \eps-transitions such that, for each
$n\ge 0$, it accepts the input~$a^{4n}$ if and only if $\cla'$
accepts~$a^n$, i.e., if and only if $a^n\in \evenodd\yes(k)$\@.
The new machine~$\cla''$ does not accept any input the length of
which is not an integer multiple of~$4$\@.

Therefore, with $11k\!+\!8$ states, $\cla''$~recognizes the
language
\bdisplay
  & \set{ (a^4)^{m\cdot 2^{k}} \st m \mbox{ is even} } =
  \set{ a^{m\cdot 2^{k+2}} \st m \mbox{ is even} } =
  \evenodd\yes(k\!+\!2) \,.
\edisplay
Using the substitution $k'=k\!+\!2$, we thus get that
$\evenodd\yes(k')$ can be recognized by a one-way \eps-free
\afa~$\cla''$ with $11(k'\!-\!2) +8 =11k'\!-\!14$ states, for each
$k'\ge 3$\@.
\eproof

To show that the gap exhibited by Theorems \ref{t:nfa-even-odd}
and~\ref{t:afa-even-odd} is tight, we need an upper bound for
converting alternation into determinism on unary promise problems:

\blemma{l:unary-afa-to-dfa}
If a promise problem $\ttp=(\ttp\yes,\ttp\no)$ built over a unary
input alphabet can be solved by a \ow\afa\ with $n$ states, it
can also be solved by a \ow\dfa\ with $t(n)\le 2^n$ states.
\elemma
\bproof
As pointed out in~\cite[Sect.~5]{MePi01}, each \ow\afa~\cla\ with
$n$~states recognizing a unary language~\ttu\ can be replaced by
an equivalent \ow\dfa~\cld\ with $2^n$~states. This is a simple
consequence of the fact that \cla~can be replaced by a
\ow\dfa~\cld\ with $2^n$ states recognizing~$\ttu\rvs$, the
reversal of the original language~\cite{CKS81}\@. (Actually, the
construction in~\cite{CKS81} works for Boolean automata\,---\,see
Footnote~\ref{ft:boolean}\,---\,which covers, as a special case,
the classical alternating automata as well\@.) However, for unary
languages, we have $\ttu\rvs=\ttu$\@.

Then, by the same reasoning as used in the proof of
Theorem~\ref{t:afa-to-dfa}, this trade-off easily extends to unary
promise problems.
\eproof

By the above lemma, the lower bound for solving $\evenodd(k)$ by
\ow\afas\ is $k\!+\!1$ states, since a \ow\afa\ with less than
$k\!+\!1$ states would give us a \ow\dfa\ with less than $2^{k+1}$
states for this promise problem, which contradicts
Theorem~\ref{t:nfa-even-odd}\@.

By combining Theorems \ref{t:nfa-even-odd}
and~\ref{t:afa-even-odd} with Lemma~\ref{l:unary-afa-to-dfa}, we
thus get:

\bcorollary{c:gap-even-odd}
The tight gap of succinctness between two-way \nfas\ and one-way
\eps-free \afas\ is asymptotically exponential on promise
problems, even for a unary input alphabet.
\ecorollary

\bsection{Las Vegas Probabilistic Finite Automata}{s:LV}
In the case of language recognition, the gap of succinctness
between Las Vegas realtime \pfas\ and one-way \dfas\ can be at
most quadratic and this gap is tight~\cite{DHRS97,HS01}\@. Quite
surprisingly, despite of this fact and despite of the fact that
the existing gaps for language recognition are not be violated on
promise problems by classical models of automata
(cf.~Corollary~\ref{c:not-violated}), we shall show here that this
does not hold for the case of one-way Las Vegas probabilistic
versus one-way deterministic automata; by switching {}from
language recognition to promise problems, the quadratic gap
increases to a tight exponential gap.

\medbreak
For this purpose, we introduce a family of promise problems
solvable by Las Vegas \ow\pfas\ with a linear number of states and
success probability arbitrarily close to~$1$, but for which any
\ow\dfas\ require an exponential number of states.%
\footnote{Remark that, in~\cite{GY14A}, we argued the same
  exponential separation by using some other families of promise
  problems, but it appeared that some of these families can also
  be solved by \ow\dfas\ with a linear number of states, and hence
  they cannot be used as witness families for proving such
  separation. More precisely, a single sentence statement given
  just before Corollary~4 in~\cite{GY14A} is not correct\@.}

Consider a family of promise problems
$\trios(n,r)$, for $n,r\in\nnumber$, such that
\bdisplay
  \trios\yes(n,r) &=& \set{\, \#x_1x_1y_1\#x_2x_2y_2\cdots \#x_rx_ry_r \st
    x_1,y_1,\ldots,x_r,y_r\in\set{0,1}^n, \\
  && \phantom{\{\,}
    \forall i\in\set{1,\ldots,r}\ \exists j\in\set{1,\ldots,n}: x_{i,j}<y_{i,j} \,} \,,\\[1.0ex]
  \trios\no(n,r) &=& \set{\, \#x_1y_1x_1\#x_2y_2x_2\cdots \#x_ry_rx_r \st
    x_1,y_1,\ldots,x_r,y_r\in\set{0,1}^n, \\
  && \phantom{\{\,}
    \forall i\in\set{1,\ldots,r}\ \exists j\in\set{1,\ldots,n}: x_{i,j}>y_{i,j} \,} \,,
\edisplay
where $x_{i,j},y_{i,j}\in\set{0,1}$ denote the $j$\xth~bit in the
respective strings $x_i,y_i\in\set{0,1}^n$\@.

\btheorem{t:lv-trios}
For each $n,r\in\nnumber$, the promise problem $\trios(n,r)$ can
be solved by a Las Vegas \ow\pfa~\clp\ with $4n\!+\!3$ states and
success probability $1-(\frac{n-1}{n})^r$\@.
\etheorem
\bproof
The automaton~\clp\ uses the following state set:
\bdisplay
  Q &=& \set{q\ini,q\acc,q\rej} \,\cup\, \set{ s_i \st i\in\set{1,\ldots,n} } \,\cup\, \\
  && \set{ t_{i,0} \st i\in\set{1,\ldots,2n} } \,\cup\, \set{ t_{i,1} \st i\in\set{1,\ldots,n} } \,,
\edisplay
where $q\ini$~is the initial and also the \quo{don't know} state,
while $q\acc,q\rej$ are the accepting and rejecting states,
respectively.

Recall that we do not have to worry about the outcome of the
computation if the input is not of the form $(\#\set{0,1}^{3n})^r$\@.
Now, let~$\#x_iu_iv_i$, with $x_i,u_i,v_i\in\set{0,1}^n$, denote
the $i$\xth~input segment. Typically, for $i>1$, the processing
of the $i$\xth~segment is activated already somewhere in the
middle of the previous segment, by switching to the initial
state~$q\ini$\@. For $i=1$, the machine begins with the first
symbol~$\#$\@. For these reasons, in the state~$q\ini$, the
machine ignores all symbols along the input until it reads the
separator~$\#$, when it enters the state~$s_1$\@. This is
implemented by transitions $q\ini\trs{b} q\ini$, for each
$b\in\set{0,1}$, together with $q\ini\trs{\#} s_1$\@. (Unless
otherwise stated explicitly, all transitions throughout this proof
are \quo{deterministic}, i.e., with probability~$1$\@.)

Next, starting in the state~$s_1$, the automaton~\clp\ moves along
the string~$x_i$ and chooses, with equal
probability~$\frac{1}{n}$, one of the symbols
$x_{i,1},\ldots,x_{i,n}\in \set{0,1}$ for subsequent verification.
This is implemented as follows. If \clp~is in the state~$s_j$, for
$j\in\set{1,\ldots,n}$, and the next input symbol is
$b\in\set{0,1}$, then \clp~reads this symbol by switching to the
state~$t_{1,b}$ with probability
$p_j= \frac{1}{p'_1\cdots p'_{j-1}}\cdot \frac{1}{n}$ \,and to the
state~$s_{j+1}$ with probability $p'_j= 1\!-\!p_j$\@. (As usual in
mathematical notation, we take $p'_1\cdots p'_{j-1}=1$ for
$j=1$\@.) The reader may easily verify that the sequence of
probabilities $p_1,\ldots,p_n$ ends with $p_n=1$, which gives
$p'_n=0$, and hence there is no need to introduce a
state~$s_{n+1}$\@. It is also easy to see that \clp~gets {}from
the state~$s_1$ at the beginning of $x_i= x_{i,1}\cdots x_{i,n}$
to the state~$t_{1,x_{i,j}}$ positioned just behind the
symbol~$x_{i,j}$ with probability~$\frac{1}{n}$\@.

The next phase depends on whether \clp~is in the state $t_{1,0}$
or in~$t_{1,1}$, that is, on whether the chosen bit $x_{i,j}$
is equal to $0$ or to~$1$\@.

For the case of $x_{i,j}=0$, the routine starting in the
state~$t_{1,0}$ is responsible to verify that $v_{i,j}=1$, which
corresponds to the case of $\#x_iu_iv_i= \#x_ix_iy_i$, with
$x_{i,j}<y_{i,j}$\@. Since \clp~gets to~$t_{1,0}$ just after
reading the symbol~$x_{i,j}$, this only requires to examine the
bit that is placed exactly $2n$~positions to the right. This is
implemented by transitions $t_{j,0}\trs{b} t_{j+1,0}$, for each
$j\in \set{1,\ldots,2n\!-\!1}$ and each $b\in\set{0,1}$\@.
Finally, if $v_{i,j}=1$, then \clp~accepts, using
$t_{2n,0}\trs{1} q\acc$\@. Otherwise, \clp~switches back to the
initial state, by $t_{2n,0}\trs{0} q\ini$\@. In the accepting
state~$q\acc$, the machine just consumes the remaining part of the
input, using $q\acc\trs{a} q\acc$, for each $a\in\set{0,1,\#}$\@.
Conversely, in the initial state~$q\ini$, the machine proceeds to
examine the next input segment~$\#x_{i+1}u_{i+1}v_{i+1}$, by
transitions already described above. However, if \clp~switches
to~$q\ini$ while processing the last input segment (for $i=r$), it
halts in~$q\ini$ at the end of the input. This is interpreted as a
\quo{don't know} answer.

The case of $x_{i,j}=1$ is very similar. This time the routine
starting in the state~$t_{1,1}$ verifies that $u_{i,j}=0$, which
corresponds to the case of $\#x_iu_iv_i= \#x_iy_ix_i$, with
$x_{i,j}>y_{i,j}$\@. This only requires to examine the bit placed
exactly $n$~positions to the right of the symbol~$x_{i,j}$\@. This
is implemented by transitions $t_{j,1}\trs{b} t_{j+1,1}$, for
each $j\in \set{1,\ldots,n\!-\!1}$ and each $b\in\set{0,1}$\@.
After that, if $u_{i,j}=0$, then \clp~rejects, using
$t_{n,1}\trs{0} q\rej$\@. Otherwise, \clp~switches back to the
initial state, by $t_{n,1}\trs{1} q\ini$\@. In the rejecting
state~$q\rej$, the machine consumes the rest of the input, using
$q\rej\trs{a} q\rej$, for each $a\in\set{0,1,\#}$\@. Conversely,
in the initial state~$q\ini$, the machine proceeds in the same way
as in the case of $x_{i,j}=0$; either it proceeds to the next
input segment, or else it halts in~$q\ini$ at the end of the
input, giving a \quo{don't know} answer.

\smallbreak
Summing up, if the $i$\xth~segment is of the form
$\#x_iu_iv_i= \#x_ix_iy_i$, with $x_{i,j}<y_{i,j}$ for some
$j\in\set{1,\ldots,n}$, then~\clp, starting to process this
segment in the state~$q\ini$, accepts the entire input with
probability at least~$\frac{1}{n}$ (the exact value increases in
the number of bit pairs satisfying $x_{i,j}<y_{i,j}$) and proceeds
to the next segment in the state~$q\ini$ with probability at most
$1\!-\!\frac{1}{n}$\@. (The machine never rejects on such segment,
since $u_i=x_i$ and there are no bit pairs satisfying
$x_{i,j}>u_{i,j}$\@.)

Now, recall that each $w\in \trios\yes(n,r)$ consists of $r$
segments having this form, and hence \clp~gives the
\quo{don't know} answer with probability at most
$(1\!-\!\frac{1}{n})^r$ for~$w$\@. Therefore, $w$~is accepted with
probability at least $1-(1\!-\!\frac{1}{n})^r$ and never rejected.

\smallbreak
Conversely, if $i$\xth~segment is of the form
$\#x_iu_iv_i= \#x_iy_ix_i$, with $x_{i,j}>y_{i,j}$ for some
$j\in\set{1,\ldots,n}$, then, in the course of processing this
segment, \clp~rejects the entire input with probability at
least~$\frac{1}{n}$ and proceeds to the next segment with
probability at most $1\!-\!\frac{1}{n}$\@. (Here the machine never
accepts, since $v_i=x_i$ and there are no bit pairs with
$x_{i,j}<v_{i,j}$\@.)

But then each $w\in \trios\no(n,r)$, consisting of $r$ segments of
this form, is rejected with probability at least
$1-(1\!-\!\frac{1}{n})^r$ and never accepted, by the same
reasoning as for $\trios\yes(n,r)$\@.
\eproof

It is obvious that, for arbitrarily small fixed constant
$\eps>0$, by taking $r= \ceil{\log\eps/\log(\frac{n-1}{n})}$, we
shall obtain a family of promise problems, namely,
$\trios'_{\eps}(n)$, for $n\in\nnumber$, where
$\trios'_{\eps}(n)= \trios(n,\ceil{\log\eps/\log(\frac{n-1}{n})})$,
such that the $n$\xth~member of this family can be solved by a Las
Vegas \ow\pfa\ with $4n\!+\!3$ states and success probability at
least \mbox{$1\!-\!\eps$}\@.

On the other hand, for each fixed $r\in\nnumber$, \ow\dfas\ need
an exponential number of states for solving $\trios(n,r)$\@.

\btheorem{t:dfa-trios}
For each $n,r\in\nnumber$, any \ow\dfa\ needs at least $2^n$
states for solving the promise problem $\trios(n,r)$\@.
\etheorem
\bproof
For contradiction, suppose that this promise problem can be solved
by a \ow\dfa~\cld\ using $\card{S}<2^n$ states. Let $s\ini$~be its
initial state, and let $\delta(q,u)\in S$ denote the state reached
by~\cld\ {}from the state $q\in S$ by reading the string
$u\in\Sigma\str$\@. Clearly,
$\delta(q,uv)= \delta(\delta(q,u),v)$, for each state~$q$ and each
two strings~$u,v$\@.

Now, for each state $q\in S$, consider the enumeration of states
$\delta(q,\#u)$, running over all strings $u\in\set{0,1}^n$\@.
Since the number of states in~$S$ is smaller than the number of
strings in $\set{0,1}^n$, we have, for each $q\in S$, that there
must exist at least two different strings $u,v\in\set{0,1}^n$ with
$\delta(q,\#u)= \delta(q,\#v)$\@. Using some lexicographic
ordering for $\set{0,1}^n$, we can pick $u_q$ and~$v_q$, the first
pair of strings satisfying $u_q\ne v_q$ and
$\delta(q,\#u_q)= \delta(q,\#v_q)$\@. Moreover, without loss of
generality, we can assume that, for some bit position
$j\in \set{1,\ldots,n}$, we have $u_{q,j}< v_{q,j}$\@. (Here
$u_{q,j},v_{q,j}$ denote the $j$\xth~bit in the respective
strings~$u_q,v_q$\@.) Otherwise, by renaming, the roles of $u_q$
and~$v_q$ can be swapped\@.

Consider now the input
\bdisplay
  w\acc &=& \#u_{q_1}u_{q_1}v_{q_1} \#u_{q_2}u_{q_2}v_{q_2}\cdots
    \#u_{q_r}u_{q_r}v_{q_r} \,, \mbox{ \ where}\\
  q_1 &=& s\ini \,,\\
  q_{i+1} &=& \delta(q_i,\#u_{q_i}u_{q_i}v_{q_i}) \,,
    \mbox{ \ for $i= 1,\ldots,r$} .
\edisplay
It should be easily seen that, along the input~$w\acc$,
\,\cld~passes the segment boundaries by the sequence of states
$q_1,q_2,\ldots,q_r,q_{r+1}$\@. More precisely, for each
$i= 1,\ldots,r$, the machine is in the state~$q_i$ at the
beginning of the $i$\xth\ segment $\#u_{q_i}u_{q_i}v_{q_i}$,
halting in the state~$q_{r+1}$ at the end of the last segment. It
is also obvious that $w\acc\in \trios\yes(n,r)$, since, for each
segment $\#u_{q_i}u_{q_i}v_{q_i}$, we have a bit position
$j\in \set{1,\ldots,n}$ satisfying $u_{q,j}< v_{q,j}$\@.
Therefore, $q_{r+1}$~is an accepting state.

Next, consider the input
\bdisplay
  w\rej &=& \#v_{q_1}u_{q_1}v_{q_1} \#v_{q_2}u_{q_2}v_{q_2}\cdots
    \#v_{q_r}u_{q_r}v_{q_r} \,.
\edisplay
Clearly, also here the computation starts in the state $q_1=s\ini$\@.
Now, using the fact that $\delta(q,\#u_q)= \delta(q,\#v_q)$ for
each state $q\in S$, we have that $q_{i+1}$ can also be expressed
as
\bdisplay
  q_{i+1} &=& \delta(q_i,\#u_{q_i}u_{q_i}v_{q_i}) =
    \delta(\delta(q_i,\#u_{q_i}),u_{q_i}v_{q_i}) \\[1.0ex]
  &=& \delta(\delta(q_i,\#v_{q_i}),u_{q_i}v_{q_i}) =
    \delta(q_i,\#v_{q_i}u_{q_i}v_{q_i}) \,,
\edisplay
for each $i= 1,\ldots,r$\@. Therefore, along the input~$w\rej$,
\,\cld~passes the segment boundaries by the same sequence of
states $q_1,q_2,\ldots,q_r,q_{r+1}$, halting in~$q_{r+1}$\@. But
then, since $q_{r+1}$~is an accepting state, \cld~accepts~$w\rej$\@.
However, $w\rej\in \trios\no(n,r)$, since, for each segment
$\#v_{q_i}u_{q_i}v_{q_i}$, we have a bit position
$j\in \set{1,\ldots,n}$ satisfying $v_{q,j}>u_{q,j}$\@.

Thus, each \ow\dfa~\cld\ with less than $2^n$ states accepts an
input that should be rejected, a contradiction.
\eproof

As a consequence of the above theorem, we get the asymptotically
optimal exponential size~$\Theta(2^n)$ for \ow\dfas\ solving
$\trios(n,r)$; the construction of a \ow\dfa\ with $O(2^n)$ states
is quite straightforward.

On the other hand, Theorem~\ref{t:lv-trios} guarantees $4n\!+\!3$
states for Las Vegas \ow\pfas\ solving $\trios(n,r)$, and hence we
have an exponential lower bound for the gap between these two
models of automata. Finally, by Corollary~\ref{c:lv-to-dfa}, we
have an exponential upper bound, which gives:

\bcorollary{c:gap-lv}
The tight gap of succinctness between one-way \dfas\ and Las Vegas
one-way \eps-free \pfas\ is asymptotically exponential on promise
problems.
\ecorollary

Due to Theorem~\ref{t:lv-trios}, we know that $\trios(n,n^2)$ can
be solved by a Las Vegas \ow\pfa\ with $4n\!+\!3$ states and
success probability $\sigma(n)\ge 1-(1\!-\!\frac{1}{n})^{n^2}\!$\@.
Thus, using the fact~\cite[Lm.~A.3.60+]{Hr05} that
$(1\!-\!\frac{1}{x})^x< \frac{1}{e}$ \,for each real $x>1$, the
probability of a \quo{don't know} outcome is at most
\bdisplay
  & 1\!-\!\sigma(n) \le \left(1\!-\!\frac{1}{n}\right)^{n\cdot n} <
  \left(\frac{1}{e}\right)^n = \frac{1}{e^n} \,.
\edisplay
This Las Vegas \ow\pfa\ can be converted to an equivalent
restarting one-way \pfa, executing the procedure in an infinite
loop, by restarting the entire computation when the original
machine halts in the \quo{don't know} state at the end of the
input. (This can be easily achieved by using a single additional
state\@.) Therefore, we can obtain a restarting \ow\pfa\ with
$4n\!+\!4$ states, solving $\trios(n,n^2)$ exactly. The expected
number of the input tape sweeps can be bounded by
\bdisplay
  & \varrho(n) =
  \sum_{i=1}^{\infty} i\!\cdot\! \sigma(n)\!\cdot\! (1\!-\!\sigma(n))^{i-1} \le
  \sum_{i=1}^{\infty} i\!\cdot\! 1\!\cdot\! \left(\frac{1}{e^n}\right)^{i-1} =
  \frac{1}{\textstyle \left(1\!-\!\frac{1}{e^n}\right)^2} =
  \left(1\!+\!\frac{1}{e^n-1}\right)^2 \le
  1\!+\!o(1) \,.
\edisplay
Thus, for the input of length~$\abs{w}$, the expected running time
is $(1\!+\!o(1))\!\cdot\!\abs{w}$\@.

\medbreak
Finally, the family of promise problems $\trios(n,r)$, for
$n,r\in\nnumber$, does not give corresponding separations for
two-way machines, since already \tw\dfas\ can solve $\trios(n,r)$
with $O(n)$ states, for each~$r$\@. This is done as follows.
Let~$\#x_iu_iv_i$, with $x_i,u_i,v_i\in\set{0,1}^n$, denote the
$i$\xth~segment along the input. Processing of this segment begins
with positioning the input head at the $n$\xth~bit of~$u_i$, by
moving $2n$ positions to the right {}from the symbol~$\#$\@. Then,
for $j=n,\ldots,1$ (in that order), the machine verifies that
$u_{i,j}=x_{i,j}$\@. (To move {}from the $j$\xth~bit of~$u_i$ to
the $j$\xth~bit of~$x_i$, just travel exactly $n$ positions to the
left; to move {}from the $j$\xth~bit of~$x_i$ to the
$(j\!-\!1)$\xst~bit of~$u_i$, just travel exactly $n\!-\!1$
positions to the right. The machine detects that it has tested all
$n$ bits {}from the fact that, after traveling exactly $n$ positions
to the left for the $j$\xth~bit of~$x_i$, it finds the symbol~$\#$
at this position, instead of zero or one\@.) In a similar way, the
machine can also search for a bit position~$j$ satisfying
$u_{i,j}<v_{i,j}$, this time iterating in ascending order, for
$j=1,\ldots,n$\@. If the $i$\xth~input segment passes the test
successfully, the machine proceeds to examine the next
segment~$\#x_{i+1}u_{i+1}v_{i+1}$ or, for $i=r$, it accepts.

\bsection{Two-Sided Bounded-Error Probabilistic Finite Automata}{s:bounded-error}
In Section~\ref{s:basics}, we have shown the limitations of Las
Vegas \pfas\@. One-sided error%
\footnote{We also cover the case of \emph{unbounded-error}, i.e.,
  the machines with no bound on the error and so the error can be
  arbitrarily close to~$1$ for some inputs\@.}
\pfas\ have similar limitations since they can be simulated by
nondeterministic or universal finite automata by removing the
probabilities: If a promise problem $\ttp=(\ttp\yes,\ttp\no)$ can
be solved by a \pfa\ rejecting each $w\in\ttp\no$ with
probability~$1$ (only members of $\ttp\yes$ are accepted), then we
obtain an \nfa\ {}from the one-sided error \pfa\@. Conversely, if
each $w\in\ttp\yes$ is accepted with probability~$1$ (only members
of $\ttp\no$ are rejected), then we obtain a finite automaton
making only universal decisions (or an \nfa\ for~$\ttp\no$) {}from
the one-sided error \pfa\@. In this section, we show that we have
a different picture for \pfas\ with two-sided bounded-error.

\medbreak
First, we show that bounded-error \ow\pfas\ can be very succinct
compared to Las Vegas \tw\pfas, \tw\afas, or any other simpler
machines. In fact, we present a parallel result to that of
Ambainis and Yakary{\i}lmaz~\cite{AY12} but with bounded error
instead of exact acceptance.

To this aim, consider~$\clu_p$, a \ow\pfa\ with the unary input
alphabet $\Sigma=\set{a}$ and two states. Namely, let $s\ini$~be
the initial and also the only accepting state, and $s\rej$~be the
rejecting state. In the state~$s\ini$, the machine reads the
symbol~$a$ staying in~$s\ini$ with probability~$p$ and switching
to~$s\rej$ with the remaining probability $p'= 1\!-\!p$\@.
In~$s\rej$, the machine just consumes the remaining part of the
input, executing a single-state loop.

Now, let $\up(p)$~be a promise problem such that
\bdisplay
  \up\yes(p) &=& \set{ a^j \st f_{\clu_p}(a^j)\ge \frac{3}{4} } \,,\\[1.0ex]
  \up\no(p)  &=& \set{ a^j \st f_{\clu_p}(a^j)\le \frac{1}{4} } \,.
\edisplay
Clearly, for each $p\in (0,1)$, the promise problem $\up(p)$ can
be solved by a bounded-error one-way \eps-free \pfa\ with only two
states, namely, by~$\clu_p$\@.

On the other hand, it is easy to show that the number of states
required by any \tw\afa\ for solving $\up(p)$ \,(which covers all
simpler models of automata as well) increases when $p$
approaches~$1$\@: Since it is straightforward that the string
$a^j$ is accepted by~$\clu_p$ with probability~$p^j$ for each
$j\ge 0$, we get $a^j\in \up\yes(p)$ only for finitely many values
satisfying $j\le \flor{\log_p\frac{3}{4}}$, whereas $\up\no(p)$
contains infinitely many strings satisfying
$j\ge \ceil{\log_p\frac{1}{4}}$\@. Let us denote these two
critical lengths as
\bdisplay
  A_p = \flor{\log_p\frac{3}{4}} &\mbox{ and }&
  R_p = \ceil{\log_p\frac{1}{4}} \,.
\edisplay
Note also that $A_p\rightarrow +\infty$ as $p\rightarrow 1$\@.

\btheorem{t:up-exact}
For one-way/two-way deterministic/nondeterministic finite
automata, and each $p\in (0,1)$, the number of states that is
sufficient and necessary for solving the promise problem $\up(p)$
is exactly $A_p\!+\!1$\@.
\etheorem
\bproof
First, we can easily design a \ow\dfa~\cld\ (hence, any more
powerful model as well) solving $\clu_p$ with $A_p\!+\!1$ states.
After the initial chain of $A_p\!+\!1$ accepting states,
responsible to accept~$a^j$ for $j\in \set{0,\ldots,A_p}$, the
automaton enters a rejecting state~$s\rej$, in which it executes a
single-step loop, consuming the rest of the input. By allowing
undefined transitions in~\cld, the state~$s\rej$ can be
eliminated.

Second, before providing the matching lower bound, we need to
present the so-called \quo{$n\rightarrow n\!+\!n!$}
method~\cite{SHL65,CHR91,Ge91} (cf.~also the proof of
Theorem~\ref{t:nfa-even-odd})\@. To make the paper self-contained,
we include also a proof, for a simplified version of this method
that is sufficient for our purposes.

\bclaim
Let \cln~be a \tw\nfa\ (or any less powerful model) with $n$
states and let $a^m$ be a unary string, with $m\ge n$\@. Then, if
\cln~has a computation path traversing across~$a^m$ {}from left to
right, starting in some state~$q_1$ and ending in some~$q_2$, it
has also a computation path traversing
across~$a^{m+h\cdot m!}$, for each $h\ge 1$, starting and ending
in the same states $q_1$ and~$q_2$\@.
\eclaim
The proof is based on the fact that \cln\ uses only $n$~states,
but the sequence of states along the left-to-right traversal
across~$a^m$ is of length $m\!+\!1>n$ (including $q_1$ at the
beginning and~$q_2$ at the end)\@. Thus, in the course of this
traversal, some state must be repeated. Hence, \cln~executes a
loop, by which it travels some $\ell\le m$ positions to the right.
Note that $m!/\ell$ is a positive integer. Hence, for each
$h\ge 1$, by using $h\!\cdot\!m!/\ell$ additional iterations of
this loop traveling $\ell$~positions, the machine traverses the
string~$a^{m+h\cdot m!}$, starting and ending in the same states
$q_1$ and~$q_2$\@. This completes the proof of the claim.

\medbreak
Now, let \cln~be a \tw\nfa\ (or any less powerful model) solving
$\up(p)$ with at most $A_p$ states. Consider the inputs $a^m$
and~$a^{m+h\cdot m!}$, for $m=A_p$ and arbitrary $h\ge 1$\@.
Clearly, $a^m= a^{A_p}\in \up\yes(p)$, and hence \cln\ has at
least one accepting computation path for this input.

However, by the above claim, whenever this computation path
traverses {}from the left endmarker of~$a^m$ to the opposite
endmarker, starting in some state~$q_1$ and ending in some~$q_2$,
an updated version of this path can traverse
across~$a^{m+h\cdot m!}$ starting and ending in the same states
$q_1$ and~$q_2$\@. By symmetry, the same holds for right-to-left
traversals. Moreover, a U-turn starting and ending at the same
endmarker of~$a^m$ can obviously do the same on the
input~$a^{m+h\cdot m!}$\@. Thus, for each $h\ge 1$, \,\cln~has a
computation path accepting~$a^{m+h\cdot m!}$\@. But then
\cln~accepts infinitely many inputs, which contradicts the fact
that all inputs longer than~$R_p$ belong to $\up\no(p)$ and should
be rejected. Thus, \cln~must use at least $A_p\!+\!1$ states.
\eproof

A~lower bound for \tw\afas\ solving the promise problem $\up(p)$
can be obtained by combining Corollary~\ref{c:not-violated},
Theorem~\ref{t:up-exact}, and the result of Geffert and
Okhotin~\cite{GO14}, stating that an arbitrary $n$-state \tw\afa\
can be converted into an equivalent \ow\nfa\ with
$2^{O(n\cdot\log n)}$ states. {}From this one can derive that the
number of states in any \tw\afa\ solving $\up(p)$ is at least
$\Omega(\frac{\log A_p}{\log\log A_p})$\@.

\bcorollary{c:gap-bounded-error}
There exists $\set{\up(p) \st p\in (0,1)}$, a family of unary
promise problems such that bounded-error one-way \eps-free \pfas\
with only two states are sufficient to solve all family, but the
number of states required by \tw\afas\ (or by any simpler
machines) to solve this family cannot be bounded by any constant.
\ecorollary

The error bound~$\frac{1}{4}$ in the definition of~$\up(p)$ can be
replaced with an arbitrarily small error; the same results will
follow by the use of the same reasoning.

\medbreak
Next, we present a separation result between bounded-error
\ow\pfas\ and \ow\dfas\ (and so \tw\afas, or any other model
capable of recognizing only regular languages)\@. For this
purpose, we use an idea given by Jibran and
Yakary{\i}lmaz~\cite{RY14A}\@. It is known that \tw\pfas\ can
recognize some nonregular languages, e.g.,
$\eq= \set{a^nb^n \st n\in\nnumber}$~\cite{Fr81}, with bounded
error but this requires exponential time~\cite{DS90}\@. It was
also shown that $\eq$ can be recognized by a restarting \ow\pfa\
for any error bound~\cite{YS10B}\@. If the given string~$a^mb^n$,
where $m,n\in\nnumber$, can be examined exponentially many times,
then the algorithm given in~\cite{YS10B} for~$\eq$ can distinguish
between the cases of $m=n$ and $m\ne n$ with high probability.
Based on this, we introduce a new family of promise problems
$\expeq(c)$, for integer $c\ge 3$\@:
\bdisplay
  \expeq\yes(c) &=& \set{ (a^mb^n)^{3\cdot(2c^2)^{m+n}\cdot\ceil{\ln c}}
    \st m,n\in\nnumber, \,m=n } \,,\\
  \expeq\no(c) &=& \set{ (a^mb^n)^{3\cdot(2c^2)^{m+n}\cdot\ceil{\ln c}}
    \st m,n\in\nnumber, \,m\ne n } \,.
\edisplay

By~\cite{YS10B}, for each integer $c\ge 3$, there exists a
restarting \ow\pfa~$\clp_c$ recognizing~$\eq$ such that, in one
pass along each given input~$a^mb^n$, the probability of giving an
erroneous decision (rejecting if the input should be accepted or
accepting if it should be rejected) is at least $c$~times smaller
than the probability of giving a correct decision. {}From this
moment on, one pass of~$\clp_c$ along the given input~$a^mb^n$
will be called a \quo{round}\@. To be more exact, let $\mba$
and~$\mbr$ denote, respectively, the probability that
$\clp_c$~accepts and rejects the input~$a^mb^n$ in a single round.
Then, by~\cite{YS10B}, the following holds:
\bequations
  \mbr &\le& \frac{1}{c}\!\cdot\!\mba \,, &\mbox{ \ if $m=n$} \,,\\[1.0ex]
  \mba &\le& \frac{1}{c}\!\cdot\!\mbr \,, &\mbox{ \ if $m\ne n$} \,.
\eequations{e:ar}
The remaining probability $\mbn= 1\!-\!\mba\!-\!\mbr$ represents
computations ending with a \quo{don't know yet} result, after
which $\clp_c$ restarts another round. In addition, we know
{}from~\cite{YS10B} that
\bequations
  \mba &=& \frac{1}{3\cdot(2c^2)^{m+n}} \,,
    & \mbox{ \ independently of whether $m=n$} \,.
\eequations{e:a}
Note also that by combining (\ref{e:a}) and~(\ref{e:ar}) we get
$\mba>0$ for $m=n$ and $\mbr>0$ for $m\ne n$\@.

Finally, one of nice properties of these restarting \ow\pfas\ is
that they all use the same number of states, which is a constant
that does not depend on~$c$\@. (This is achieved by using
arbitrarily small transition probabilities\@.)

Now we are ready to construct a \ow\pfa~$\clp'_c$ for solving
$\expeq(c)$\@. For the given input~$(a^mb^n)^t$, where
\bequations
  t &=& 3\!\cdot\!(2c^2)^{m+n}\!\cdot\!\ceil{\ln c} \,,
\eequations{e:t}
$\clp'_c$ simulates~$\clp_c$ on~$a^mb^n$ but, each time
$\clp_c$~restarts its computation {}from the very beginning
of~$a^mb^n$, the machine~$\clp'_c$ proceeds to the next copy
of~$a^mb^n$ along its own input. More precisely, if $\clp_c$~gives
the decision of \quo{acceptance} in the course of one round,
$\clp'_c$~immediately accepts, by switching to an accepting
state~$s\acc$\@. Similarly, if $\clp_c$~gives the decision of
\quo{rejection}, $\clp'_c$~immediately rejects, switching to a
rejecting~$s\rej$\@. (In both these states, $\clp'_c$~consumes the
rest of the input by executing single-state loops\@.) Otherwise,
$\clp'_c$~gets to the end of the current copy of~$a^mb^n$ in a
state~$s\neu$, representing here a \quo{don't know yet} outcome.
That is, $\clp'_c$~is ready to retry the entire procedure with the
next copy of~$a^mb^n$\@. Clearly, along the input~$(a^mb^n)^t$,
this can be repeated $t$~times, after which $\clp'_c$~halts in the
state~$s\neu$\@.

Since the number of states in~$\clp_c$ is a constant not depending
on~$c$, it is obvious that $\clp'_c$~also uses a constant number
of states that does not depend on~$c$\@.

Now, let $\mba_t$ and~$\mbr_t$ denote the respective probabilities
that $\clp'_c$~accepts and rejects the input~$(a^mb^n)^t$, by
halting in the respective state $s\acc$ or~$s\rej$\@. The
remaining probability $\mbn_t= 1\!-\!\mba_t\!-\!\mbr_t$ represents
computations not deciding about acceptance or rejection, halting
in the state~$s\neu$\@.

Let us estimate~$\mbn_t$ first. By combining $\mbr\ge 0$ with
(\ref{e:a}), (\ref{e:t}), and the fact~\cite[Lm.~A.3.60+]{Hr05}
that $(1\!-\!\frac{1}{x})^x< \frac{1}{e}$ \,for each real $x>1$
(in that order), we get, independently of whether $m=n$, that
\bdisplay
  \mbn_t &=& \mbn^t =
    (1\!-\!\mba\!-\!\mbr)^t \le
    (1\!-\!\mba)^t =
    \left( 1\!-\!\frac{1}{3\cdot(2c^2)^{m+n}} \right)^t =
    \left( 1\!-\!\frac{1}{3\cdot(2c^2)^{m+n}} \right)^{ 3\cdot(2c^2)^{m+n}\cdot\ceil{\ln c} } \\
  &<& \left( \frac{1}{e} \right)^{\ceil{\ln c}} \le
    \frac{1}{c} \,.
\edisplay
Therefore, using~(\ref{e:ar}) and $\mba>0$, we can derive that the
input~$(a^mb^n)^t$ satisfying $m=n$ is accepted with probability
at least
\bdisplay
  \mba_t &=& \sum_{i=1}^t \mba\!\cdot\!\mbn^{i-1} =
    \mba\!\cdot\!\frac{1-\mbn^t}{1-\mbn} =
    \frac{\mba}{\mba+\mbr}\!\cdot\!(1\!-\!\mbn^t) \ge
    \frac{\mba}{\mba\,+\,\mba/c}\!\cdot\!(1\!-\!\mbn^t) =
    \frac{c}{c+1}\!\cdot\!(1\!-\!\mbn^t) \\[1.0ex]
  &>& \frac{c}{c+1}\!\cdot\!(1\!-\!\frac{1}{c}) =
    1\!-\!\frac{2}{c+1} \,.
\edisplay
Similarly, using~(\ref{e:ar}) and $\mbr>0$, the input~$(a^mb^n)^t$
satisfying $m\ne n$ is rejected with probability at least
\bdisplay
  \mbr_t &=& \sum_{i=1}^t \mbr\!\cdot\!\mbn^{i-1} =
    \mbr\!\cdot\!\frac{1-\mbn^t}{1-\mbn} =
    \frac{\mbr}{\mba+\mbr}\!\cdot\!(1\!-\!\mbn^t) \ge
    \frac{\mbr}{\mbr/c\,+\,\mbr}\!\cdot\!(1\!-\!\mbn^t) =
    \frac{c}{c+1}\!\cdot\!(1\!-\!\mbn^t) \\[1.0ex]
  &>& \frac{c}{c+1}\!\cdot\!(1\!-\!\frac{1}{c}) =
    1\!-\!\frac{2}{c+1} \,.
\edisplay

Summing up, the success probability is always above
$1\!-\!\frac{2}{c+1}$ and consequently the error probability is
always below~$\frac{2}{c+1}$\@. (It should be pointed out that the
standard \pfas\ do not use \quo{don't know} states. However, we
can declare, by definition, the state~$s\neu$ to be a rejecting
state. This may potentially change all \quo{don't know} answers to
errors, rejecting inputs that should be accepted. However, this
does not change the fact that $\mba_t> 1\!-\!\frac{2}{c+1}$ for
$m=n$, nor does it decrease~$\mbr_t$ for $m\ne n$\@.)

By taking $c= \max\set{3,\ceil{\frac{2}{\eps}}\!-\!1}$ for
arbitrarily small but fixed $\eps>0$, we obtain the error
probability $\frac{2}{c+1}\le \eps$, keeping the same constant
number of states for each~\eps\@. This gives:

\btheorem{t:gap-nonregular}
For each fixed $\eps>0$, there exists a promise problem solvable
by a \ow\pfa\ with bounded error~\eps, using a constant number of
states that does not depend on~\eps, but there is no \ow\dfa\
(hence, no other machine capable of recognizing only regular
languages) solving the same problem.
\etheorem
\bproof
We only need to show that $\expeq(c)$ cannot be solved by any
\ow\dfa, for no \mbox{$c\ge 3$}\@. For contradiction, let \cld\ be
a \ow\dfa\ solving $\expeq(c)$, for some \mbox{$c\ge 3$}\@.
Without loss of generality, we assume that \cld\ does not have
undefined transitions, and hence it always halts at the end of the
input. (Otherwise, we can define all missing transitions by
switching to a single new rejecting state, in which \cld~scans the
rest of the input\@.) Let $n\ge 1$ denote the number of states
in~\cld\@.

Consider now the unary string~$a^n$\@. Using the Claim presented
in the proof of Theorem~\ref{t:up-exact}, we see that if \cld\
traverses across~$a^n$, starting in some state~$q_1$ and ending in
some~$q_2$, it will do the same also on the
string~$a^{n+h\cdot n!}$, for each $h\ge 1$\@. Clearly, the same
holds for traversals of~$b^n$ and~$b^{n+h\cdot n!}$\@.

Next, consider the input~$(a^nb^n)^t$, where
$t =3\!\cdot\!(2c^2)^{2n+2n!}\!\cdot\!\ceil{\ln c}$\@. If, on this
input, \cld~halts in a state~$q'$, then, by the observation above,
\cld~must halt in the same state~$q'$ also on the inputs
$(a^{n+n!}b^{n+n!})^t$ and~$(a^nb^{n+2n!})^t$\@. Therefore, \cld\
accepts $(a^{n+n!}b^{n+n!})^t$ if and only if it
accepts~$(a^nb^{n+2n!})^t$\@. But this is a contradiction, since
the former string should be accepted by~\cld\ while the latter
should be rejected. Consequently, there is no \ow\dfa\ solving
$\expeq(c)$\@.
\eproof

\bsection{Final Remarks}{s:final}
A~thorough study of promise problems can reveal several
interesting properties of computational models and give new
fundamental insights about them. In automata theory, promise
problems have mainly been used to show how quantum models can do
much better than the classical ones when compared to the case of
language recognition. In this paper, we initiated a systematic
work on promise problems for classical one-way finite automata
(deterministic, nondeterministic, alternating, and
probabilistic)\@. In this context, we have also shown that
randomness can do much better than other classical resources.

Promise problems can further be investigated for different
computational models and {}from different perspectives. Two-way
finite state machines and counter or pushdown automata models are
the first ones coming to the mind. Moreover, we believe that some
long-standing open problems, formulated for language recognition,
might be solved more easily in the case of promise problems.
\\

\noindent
\textbf{Acknowledgements.}
The authors thank Beatrice Palano for providing us a copy of~\cite{BMP14B}\@.

\bibliographystyle{plain}
\bibliography{tcs6}%

\newcommand{\DD}{\v{D}} \providecommand{\softl}{{l\kern-0.18em'\kern-0.04em}}
\begin{thebibliography}{10}

\bibitem{AGKY14A}
Farid Ablayev, Aida Gainutdinova, Kamil Khadiev, and Abuzer Yakary{\i}lmaz.
\newblock Very narrow quantum {OBDDs} and width hierarchies for classical
  {OBDDs}.
\newblock In {\em Descriptional Complexity of Formal Systems}, volume 8614 of
  {\em Lecture Notes in Computer Science}, pages 53--64. Springer-Verlag, 2014.

\bibitem{AY12}
Andris Ambainis and Abuzer Yakary{\i}lmaz.
\newblock Superiority of exact quantum automata for promise problems.
\newblock {\em Information Processing Letters}, 112(7):289--291, 2012.

\bibitem{BMP14B}
Maria~Paola Bianchi, Carlo Mereghetti, and Beatrice Palano.
\newblock Complexity of promise problems on classical and quantum automata.
\newblock To appear.

\bibitem{Buk80}
R.~G. Bukharaev.
\newblock Probabilistic automata.
\newblock {\em Journal of Mathematical Sciences}, 13(3):359--386, 1980.

\bibitem{CKS81}
Ashok~K. Chandra, Dexter~C. Kozen, and Larry~J. Stockmeyer.
\newblock Alternation.
\newblock {\em Journal of the ACM}, 28(1):114--133, 1981.

\bibitem{CHR91}
Richard Chang, Juris Hartmanis, and Desh Ranjan.
\newblock Space bounded computations: {Review} and new separation results.
\newblock {\em Theoretical Computer Science}, 80:289--302, 1991.

\bibitem{CL89}
Anne Condon and Richard~J. Lipton.
\newblock On the complexity of space bounded interactive proofs (extended
  abstract).
\newblock In {\em FOCS'89\@: Proceedings of the 30th Annual Symposium on
  Foundations of Computer Science}, pages 462--467, 1989.

\bibitem{DHRS97}
Pavol \DD{}uri\v{s}, Juraj Hromkovi\v{c}, Jos{\'e}~{D.\,P.} Rolim, and Georg
  Schnitger.
\newblock {Las} {Vegas} versus determinism for one-way communication
  complexity, finite automata, and polynomial-time computations.
\newblock In {\em Symposium on Theoretical Aspects of Computer Science}, volume
  1200 of {\em Lecture Notes in Computer Science}, pages 117--128.
  Springer-Verlag, 1997.

\bibitem{DS92}
Cynthia Dwork and Larry Stockmeyer.
\newblock Finite state verifiers~{I}: {The} power of interaction.
\newblock {\em Journal of the ACM}, 39(4):800--828, 1992.

\bibitem{DS90}
Cynthia Dwork and Larry~J. Stockmeyer.
\newblock A~time complexity gap for two-way probabilistic finite-state
  automata.
\newblock {\em SIAM Journal on Computing}, 19(6):1011--1123, 1990.

\bibitem{Fr81}
R\={u}si\c{n}\v{s} Freivalds.
\newblock Probabilistic two-way machines.
\newblock In {\em Mathematical Foundations of Computer Science}, volume 118 of
  {\em Lecture Notes in Computer Science}, pages 33--45. Springer-Verlag, 1981.

\bibitem{Ge91}
Viliam Geffert.
\newblock Nondeterministic computations in sublogarithmic space and space
  constructibility.
\newblock {\em SIAM Journal on Computing}, 20:484--498, 1991.

\bibitem{Ge12}
Viliam Geffert.
\newblock An~alternating hierarchy for finite automata.
\newblock {\em Theoretical Computer Science}, 445:1--24, 2012.

\bibitem{GO14}
Viliam Geffert and Alexander Okhotin.
\newblock Transforming two-way alternating finite automata to one-way
  nondeterministic automata.
\newblock In {\em Mathematical Foundations of Computer Science, {Part~I}},
  volume 8634 of {\em Lecture Notes in Computer Science}, pages 291--302.
  Springer-Verlag, 2014.

\bibitem{GY14A}
Viliam Geffert and Abuzer Yakary{\i}lmaz.
\newblock Classical automata on promise problems.
\newblock In {\em Descriptional Complexity of Formal Systems}, volume 8614 of
  {\em Lecture Notes in Computer Science}, pages 126--137. Springer-Verlag,
  2014.

\bibitem{Gol06A}
Oded Goldreich.
\newblock On promise problems: {A}~survey.
\newblock In {\em Essays in Memory of Shimon Even}, volume 3895 of {\em Lecture
  Notes in Computer Science}, pages 254--290. Springer-Verlag, 2006.

\bibitem{GQZ14A}
Jozef Gruska, Daowen Qiu, and Shenggen Zheng.
\newblock Generalizations of the distributed {Deutsch-Jozsa} promise problem.
\newblock Technical report, 2014.
\newblock arXiv:1402.7254.

\bibitem{GQZ14B}
Jozef Gruska, Daowen Qiu, and Shenggen Zheng.
\newblock Potential of quantum finite automata with exact acceptance.
\newblock Technical Report arXiv:1404.1689, 2014.

\bibitem{HMU07}
John~E. Hopcroft, Rajeev Motwani, and Jeffrey~D. Ullman.
\newblock {\em Introduction to Automata Theory, Languages, and Computation}.
\newblock Prentice Hall, 2007.

\bibitem{Hr05}
Juraj Hromkovi\v{c}.
\newblock {\em Design and Analysis of Randomized Algorithms (Introduction to
  Design Paradigms)}.
\newblock Springer-Verlag, 2005.

\bibitem{HS01}
Juraj Hromkovi\v{c} and Georg Schnitger.
\newblock On the power of {Las} {Vegas} for one-way communication complexity,
  {OBDDs}, and finite automata.
\newblock {\em Information and Computation}, 169(2):284--296, 2001.

\bibitem{JP11}
Galina Jir\'{a}skov\'{a} and Giovanni Pighizzini.
\newblock Optimal simulation of self-verifying automata by deterministic
  automata.
\newblock {\em Information and Computation}, 209:528--535, 2011.

\bibitem{Ka05m}
Christos~A. Kapoutsis.
\newblock Removing bidirectionality {}from nondeterministic finite automata.
\newblock In {\em Mathematical Foundations of Computer Science}, volume 3618 of
  {\em Lecture Notes in Computer Science}, pages 544--555. Springer-Verlag,
  2005.

\bibitem{KKM12}
Christos~A. Kapoutsis, Richard Kr\'{a}lovi\v{c}, and Tobias M\"{o}mke.
\newblock Size complexity of rotating and sweeping automata.
\newblock {\em Journal of Computer and System Sciences}, 78(2):537--558, 2012.

\bibitem{Ku73}
{Yu.\,I.} Kuklin.
\newblock Two-way probabilistic automata.
\newblock {\em Avtomatika i vy\v{c}istite{\softl}naja tekhnika}, 5:35--36,
  1973.
\newblock (Russian).

\bibitem{Ku10}
Rajendra Kumar.
\newblock {\em Theory of Automata, Languages, and Computation}.
\newblock Tata McGraw-Hill, 2010.

\bibitem{LLS84}
Richard~E. Ladner, Richard~J. Lipton, and Larry~J. Stockmeyer.
\newblock Alternating pushdown and stack automata.
\newblock {\em SIAM Journal on Computing}, 13(1):135--155, 1984.

\bibitem{MePi01}
Carlo Mereghetti and Giovanni Pighizzini.
\newblock Optimal simulations between unary automata.
\newblock {\em SIAM Journal on Computing}, 30:1976--1992, 2001.

\bibitem{MNYW05}
Yumiko Murakami, Masaki Nakanishi, Shigeru Yamashita, and Katsumasa Watanabe.
\newblock Quantum versus classical pushdown automata in exact computation.
\newblock {\em IPSJ Digital Courier}, 1:426--435, 2005.

\bibitem{Nak14A}
Masaki Nakanishi.
\newblock Quantum pushdown automata with a garbage tape.
\newblock Technical Report arXiv:1402.3449, 2014.

\bibitem{Pa71}
Azaria Paz.
\newblock {\em Introduction to Probabilistic Automata}.
\newblock Academic Press, New York, 1971.

\bibitem{Ra63}
Michael~O. Rabin.
\newblock Probabilistic automata.
\newblock {\em Information and Control}, 6:230--243, 1963.

\bibitem{RY14A}
Jibran Rashid and Abuzer Yakary{\i}lmaz.
\newblock Implications of quantum automata for contextuality.
\newblock In {\em Conference on Implementation and Application of Automata},
  volume 8587 of {\em Lecture Notes in Computer Science}, pages 318--331.
  Springer-Verlag, 2014.
\newblock arXiv:1404.2761.

\bibitem{RY14B}
Klaus Reinhardt and Abuzer Yakaryilmaz.
\newblock The minimum amount of useful space: New results and new directions.
\newblock In {\em Developments in Language Theory}, volume 8633 of {\em Lecture
  Notes in Computer Science}, pages 315--326. Springer-Verlag, 2014.

\bibitem{Sip80}
Michael Sipser.
\newblock Lower bounds on the size of sweeping automata.
\newblock {\em Journal of Computer and System Sciences}, 21(2):195--202, 1980.

\bibitem{SHL65}
Richard~Edwin Stearns, Juris Hartmanis, and Philip~M. {Lewis~II}.
\newblock Hierarchies of memory limited computations.
\newblock In {\em IEEE Conference Record on Switching Circuit Theory and
  Logical Design}, pages 179--190, 1965.

\bibitem{Wat09A}
John Watrous.
\newblock Quantum computational complexity.
\newblock In Robert~A. Meyers, editor, {\em Encyclopedia of Complexity and
  Systems Science}, pages 7174--7201. Springer-Verlag, 2009.

\bibitem{YS10B}
Abuzer Yakary{\i}lmaz and {A.\,C.}~Cem Say.
\newblock Succinctness of two-way probabilistic and quantum finite automata.
\newblock {\em Discrete Mathematics and Theoretical Computer Science},
  12(2):19--40, 2010.

\bibitem{ZGQ14A}
Shenggen Zheng, Jozef Gruska, and Daowen Qiu.
\newblock On the state complexity of semi-quantum finite automata.
\newblock In {\em Language and Automata Theory and Applications}, volume 8370
  of {\em Lecture Notes in Computer Science}, pages 601--612. Springer-Verlag,
  2014.

\bibitem{ZQGLM13}
Shenggen Zheng, Daowen Qiu, Jozef Gruska, Lvzhou Li, and Paulo Mateus.
\newblock State succinctness of two-way finite automata with quantum and
  classical states.
\newblock {\em Theorerical Computer Science}, 499:98--112, 2013.

\end{thebibliography}
\end{document}